\documentstyle[11pt]{article}

\def\theequation{\arabic{section}.\arabic{equation}}

\renewcommand{\theequation}{\thesection.\arabic{equation}}

\global\arraycolsep=1pt
\begin{document}

\hfill  BRX-TH-420, CPTH-S.553.0897

\hfill HUTP-97/A037, MIT-CTP-2666

\hfill hep-th/nnnmmyy

\hfill August, 1997

\vspace{20pt}

\begin{center}
{\large {\bf NONPERTURBATIVE FORMULAS FOR  CENTRAL FUNCTIONS\\[0pt]
OF SUPERSYMMETRIC GAUGE THEORIES}}
\end{center}

\vspace{6pt}

\begin{center}
{\sl D. Anselmi}

{\it Centre de Physique Theorique, Ecole Polytechnique, F-91128 Palaiseau
Cedex, FRANCE}
\end{center}

\vspace{6pt}

\begin{center}
{\sl D.Z. Freedman}

{\it Department of Mathematics and Center for Theoretical Physics,
Massachusetts Institute of Technology, Cambridge MA 02139, USA}
\end{center}

\vspace{6pt}

\begin{center}
{\sl M.T. Grisaru}

{\it Physics Department, Brandeis University, Waltham MA\ 02254, USA}
\end{center}

\vspace{6pt}

\begin{center}
{\sl A.A. Johansen}

{\it Lyman Laboratory, Harvard University, Cambridge, MA 02138, USA}
\end{center}

\vspace{8pt}

\begin{center}
{\bf Abstract}
\end{center}

\vspace{4pt} For quantum field theories that flow between ultraviolet and
infrared fixed points, central functions, defined from two-point
correlators
of the stress tensor and conserved currents, interpolate between central
charges of the UV and IR critical theories. We develop techniques
that allow
one to calculate the flows of the central charges and that of the Euler
trace anomaly coefficient in a general N=1 supersymmetric gauge theory.
Exact, explicit formulas for $SU(N_c)$ gauge theories in the conformal
window are given
and analysed. The Euler anomaly coefficient always satisfies the
inequality $%
a_{UV}-a_{IR}>0$. This is new evidence in strongly coupled theories that
this quantity satisfies a four-dimensional analogue of the
$c$-theorem, supporting the idea of irreversibility of the RG flow. Various
other
implications are discussed.

\vfill\eject

\section{Introduction}

In two dimensions there are many known examples of quantum field theories
that flow under the renormalization group (RG) from a conformal
fixed point
in the ultraviolet to another fixed point in the infrared. Often, it is
possible to work out exact results, using properties and techniques
that are
special to two dimensions. The operator product expansions of two stress
tensors or two conserved currents contain central charges $c$ and
$k$ which
encode fundamental properties of the conformal theories that appear
at the
limits of the flow. The Zamolodchikov $c$-theorem \cite{zamolo}
states the
important inequalities $c_{UV}-c_{IR}>0$ and $k_{UV}-k_{IR}>0$ that place
constraints on the flow and have a useful physical interpretation
in terms of RG irreversibility
and the thinning out of degrees of freedom as one moves to longer
distances.

In four dimensions it is more difficult to establish the existence of
conformal fixed points. When they exist, a quantum field theory can be
described as a radiative interpolation between pairs of  four-dimensional
conformal field theories. The problem is then to identify relevant
physical
quantities and study their renormalization group flow from one
fixed point
to the other. We call this problem the RG interpolation. While the RG
interpolation seems to be very difficult in the general case, it
simplifies
considerably for supersymmetric theories, where many examples of
interacting
conformal fixed points have been studied. In particular, supersymmetric
gauge theories in the ``conformal window'' \cite{emduality} have
nontrivial
IR fixed points; this is also the subset of theories with
electric-magnetic
duality. The general relation between the trace anomaly and the chiral
anomaly of the $R$-current in supersymmetry allows us to solve the RG
interpolation problem for supersymmetric gauge theories, and the
principal
application of these techniques is to theories in the conformal window.

Another difficulty of four dimensions is that no analogue of the
$c$-theorem
has been proved, although the coefficient $a$ of the Euler density in the
curved space trace anomaly has been proposed \cite{cardy} as a
$c$-theorem
candidate, and the flow $a_{UV}-a_{IR}$ is known to be positive in
all cases
where it can be tested. We shall have more to say about this later.

In two-dimensional conformal theories the operator product
expansions
of two stress tensors $T_{\mu\nu}$ or conserved currents $J_{\mu}$ are
closed, namely no new operators appear, but the situation is more
complicated in
higher dimensions. As in two dimensions, the $TT$ and $JJ$ OPE's define
primary central charges $c$ and $b$, respectively. But these OPE's
are not
closed \cite{mix,noialtri}. New operators $\Sigma$ with anomalous
dimension
appear, and the OPE's of the new operators define secondary central
charges $%
c^{\prime}$ and $b^{\prime}$ \cite{noialtri}. A $\Sigma$ operator can be
used to deform the theory off-criticality and its anomalous dimension $h$
then coincides with the critical value of the slope of the
$\beta$-function
\cite{noialtri}. This property indicates that  information about the
off-critical theory can be obtained by studying its critical limit.

It is reasonable to suggest that the central charges $c$,
$c^{\prime}$, $b$,
$b^{\prime}$, the Euler coefficient $a$ and the anomalous dimension
$h$ are
the fundamental parameters of four-dimensional conformal theories and to
study these quantities, which depend on the dimensionless couplings
of the
theory. The lowest order, two-loop radiative
corrections to the
central charge $c$ were obtained by several authors and in particular by
Jack \cite{jack} in the most general renormalizable theory of scalar,
spinor, and gauge fields. In \cite{noi} the secondary central charge $%
c^{\prime}$ was computed to two-loop order for general $N=1$ SUSY gauge
theories and Jack's result for $c$ was specialized to the case of
supersymmetric couplings. The general $N=1$ theory contains the
extended $%
N=4 $ theory and the $N=2$ theory with the critical number of
hypermultiplets as special cases, and it was observed that $c$ and $%
c^{\prime}$ are constant on the well-known marginal fixed lines of those
theories. It is a consequence of the $c$-theorem in two dimensions
that the
central charge $c$ is constant along lines of marginal deformation. This
suggests  that the central charges are invariants of
superconformal
field theory in four dimensions (SCFT$_4$) - they do not change within
families of continuously connected theories. The anomalous dimension $h$
does vary along marginal lines.

In ref. \cite{ccfis} non-perturbative definitions of primary and
secondary
central functions were given in terms of two- and four-point
correlators of
the stress tensor and conserved currents. The central functions
interpolate
between the critical values of the central charges. One goal of the RG
interpolation is to work out nonperturbative expressions for these
functions. We are going to show that this can be achieved in
supersymmetric
theories for all  primary central functions, as well as a special
subclass of secondary central functions.

In particular, the major results of our analysis are non-perturbative
formulas for the flow of the primary central functions $b$, $c$ and $a$,
that agree with  perturbative calculations and can be applied
within the
conformal window of $SU(N_c)$ theories with $N_f$ flavors to give exact
formulas for the total flows $b_{UV} - b_{IR}$, $c_{UV} - c_{IR}$ and $%
a_{UV} - a_{IR}$ in terms of $N_c$ and $N_f$ and no other parameters. The
first of these is strictly negative so that no $b$-theorem holds,
while the
second changes sign from positive to negative as $N_f$ increases from $%
3N_c/2 $ to $3N_c$. On the other hand, the flow of $a$ is positive in the
entire
conformal window. A byproduct of our general analysis is the marginal
constancy of the primary central functions to all orders in perturbation
theory.

At the perturbative level, while the two-loop contribution to the primary
flavor central charge $b$ turns out to vanish along marginal lines, in
agreement with the nonperturbative formulas, the radiative corrections to
the secondary flavor central charge $b^{\prime }$ are not marginally
constant. At the moment, we do not have a general nonperturbative
treatment
of secondary central functions. We hope to discuss our two-loop
calculations, which are based on an interesting application of conformal
symmetry to calculations of Feynman diagrams, and the situation of
secondary
central charges, in a subsequent paper \cite{new}.

The flow of the flavor central charge $b$
is analysed by two methods, one in Section 2
and the other in Section 3.
The second method is more direct, but only the first
can be easily extended to the gravitational central charges
$c$ and $a$ as described in Section 4.
In Section 5, our formulae for the total flows
$b_{UV}-b_{IR}$, $c_{UV}-c_{IR}$ and $a_{UV}-a_{IR}$
are discussed, and we give there our conclusions and outlook.
In Appendix A, the component gravitational anomalies are
obtained from  their curved superspace counterparts.

\newpage

\section{Flow of the flavor central charge}

\setcounter{equation}{0}

We consider $SU(N_{c})$ supersymmetric QCD with $N_{f}$ quark
flavors. The
theory contains gauge superfields $V^{a}(x,\theta
,\bar{\theta}),a=1,\ldots
,N_{c}^{2}-1$, whose physical components are the gauge potentials $A_{\mu
}^{a}(x)$ and Majorana gauginos $\lambda ^{a}(x)$. There are also chiral
quark and anti-quark superfields, $Q^{\alpha i}(x,\theta )$ and
$\widetilde{%
Q }_{\alpha i}(x,\theta )$, respectively, where $\alpha =1,\ldots ,N_{c}$
and $i=1,\ldots ,N_{f}$. The matter components of $Q^{\alpha i}$ are the
complex scalars $\phi ^{\alpha i}$ and Majorana spinors $\psi
^{\alpha i}$,
while $\widetilde{ Q}_{\alpha i}$ contains $\tilde{\phi}_{\alpha
i}$ and $%
\tilde{\psi}_{\alpha i}$. This electric theory has the usual gauge
interactions and no superpotential. Later we will extend the treatment to
the magnetic theory where there are additional gauge neutral chiral
superfields and a cubic superpotential.

The theory has an anomaly-free $SU(N_{f})_{Q}\times
SU(N_{f})_{\widetilde{Q}
}\times U(1)_{B}$ global symmetry group, and there are conserved currents
that appear as the $\theta \bar{\theta}$ components of the
superfields $%
\bar{Q}t^{A}Q$,
$\widetilde{\bar{Q}}\widetilde{t}^{A}\widetilde{Q}$, and
$B=(Q\bar{Q}-\widetilde{Q}\widetilde{\bar{Q}})/N_{c}$.
Here $t^{A}$ and
$\widetilde{t}^{A} $ are matrix generators of the fundamental and
anti-fundamental
representations of $SU(N_f)$, respectively. The component Noether
currents
of the Euclidean signature theory are (see \cite{noi} for details of the
notation)
\begin{eqnarray}
J_{\mu }^{A} &=&-\bar{\psi}\gamma _{\mu }Lt^{A}\psi +\bar{\phi}\stackrel{
\leftrightarrow }{D_{\mu }}t^{A}\phi ,  \label{eq:2.1} \\[2ex]
\widetilde{J}_{\mu }^{A} &=&-\widetilde{\bar{\psi}}\gamma _{\mu
}L\widetilde{
t}^{A}\widetilde{\psi
}+\widetilde{\bar{\phi}}\stackrel{\leftrightarrow }{
D_{\mu }}\widetilde{t}^{A}\widetilde{\phi },  \nonumber \\[2ex]
J_{\mu } &=&\frac{1}{N_{c}}\left[ \frac{1}{2}\bar{\psi}\gamma _{\mu }
\gamma_5\psi -
\frac{1}{2}\widetilde{\bar{\psi}}\gamma _{\mu }\gamma_5
\widetilde{\psi }+\bar{\phi}\stackrel{
\leftrightarrow }{D_{\mu }}\phi -\widetilde{\bar{\phi}}\stackrel{
\leftrightarrow }{D_{\mu }}\widetilde{\phi }\right] .  \nonumber
\end{eqnarray}
There are also classically conserved but anomalous Konishi and $R$ currents
that are given by
\begin{eqnarray}
K_{\mu } &=&\left[ \frac{1}{2}\bar{\psi}\gamma _{\mu }\gamma
_{5}\psi +\frac{
1}{2}\widetilde{\bar{\psi}}\gamma _{\mu }\gamma _{5}\widetilde{\psi
}+\bar{
\phi}\stackrel{\leftrightarrow }{D_{\mu }}\phi +\widetilde{\bar{\phi}}%
\stackrel{\leftrightarrow }{D_{\mu} }\widetilde{\phi }\right] ,
\label{eq:2.2} \\[2ex]
R_{\mu } &=&\frac{1}{2}\bar{\lambda}^{a}\gamma _{\mu }\gamma _{5}\lambda
^{a}-\frac{1}{6}(\bar{\psi}\gamma _{\mu } \gamma _{5}\psi
+\widetilde{\bar{
\psi}}\gamma _{\mu }\gamma _{5}\widetilde{\psi}
)+\frac{2}{3}(\bar{\phi}\stackrel{
\leftrightarrow }{D_{\mu }}\phi +\widetilde{\bar{\phi}}\stackrel{
\leftrightarrow }{D_{\mu }}\widetilde{\phi }).  \nonumber
\end{eqnarray}
The first is the $\theta\bar{\theta}$ component of
$K=Q\bar{Q}+\widetilde{Q}\widetilde{%
\bar{Q}} $, while the second is the lowest component of the supercurrent
superfield $J_{\alpha \dot{\alpha}}$ that also contains the stress tensor
and supersymmetry currents. It is well known that $K_{\mu }$ and
$R_{\mu }$
have internal anomalies \footnote{We distinguish between internal
anomalies,
involving the quantum gauge field, and external anomalies, involving an
external classical
gauge field.}, expressed by the operator equations (see \cite
{kogan})
\begin{equation}
\partial_{\mu} R^{\mu}= {\frac{1}{48\pi ^{2}}}[3N_{c}-N_{f}(1-\gamma
)]F_{\mu \nu }^{a} \tilde{F}_{\mu \nu }^{a},~~~~~~~\partial_{\mu}
K^{\mu}= {%
\frac{N_f}{16\pi ^{2}}}F_{\mu \nu }^{a}\tilde{F}_{\mu \nu }^{a}.
\end{equation}
The anomaly-free, RG invariant, combination \cite{kogan} of $K_{\mu}$ and
$R_{\mu}$, namely
\begin{equation}
S_{\mu }=R_{\mu }+\frac{1}{3}\left( 1-\frac{3N_{c}}{N_{f}}-\gamma \right)
K_{\mu },  \label{eq:2.4}
\end{equation}
will be important for us. The coefficient of $K_{\mu }$ is the
numerator of
the exact NSVZ \cite{nsvz}  $\beta$ -function
\begin{equation}
\beta (g)=-\frac{g^{3}}{16\pi ^{2}}\frac{3N_{c}-N_{f} (1-\gamma (g ))}{
1-g^{2}N_{c}/8\pi ^{2}}  \label{eq:2.5}
\end{equation}
and $\gamma/2 $ is the anomalous dimension of the superfield $Q$ (or $%
\widetilde{Q}$).

We now let $J_{\mu }(x)$ denote any one of the conserved flavor
currents of (
\ref{eq:2.1}). Its two-point current correlation function has the form
\begin{equation}
\langle J_{\mu }(x)J_{\nu }(0)\rangle =\frac{1}{16\pi
^{4}}(\partial _{\mu
}\partial _{\nu }-{\,\lower0.9pt%
\vbox{\hrule \hbox{\vrule height 0.2 cm \hskip 0.2 cm
\vrule
height 0.2 cm}\hrule}\,}~\delta _{\mu \nu }){}~\left(
{}~{\frac{b(g(1/x))}{
x^{4}}}\right) ,  \label{jj}
\end{equation}
Since $J_{\mu }(x)$ is conserved and thus has no anomalous dimension, the
Callan- Symanzik equation requires that $b(g(1/x))$ depend only on the
running coupling $g(1/x)$ that satisfies
\begin{equation}
\mu \frac{{\rm d}g(\mu )}{{\rm d}\mu }=\beta (g(\mu )).  \label{eq:2.7}
\end{equation}
As discussed in \cite{ccfis} $b(g(1/x))$ is a primary central
function that
interpolates between flavor central charges at the fixed points
$g_{UV}$ and
$g_{IR}$ of the renormalization group flow. Specifically
\begin{eqnarray}
b_{UV} &=&\lim_{x\longrightarrow 0}b(g(1/x))=b(g_{UV})  \label{eq:2.8} \\
b_{IR} &=&\lim_{x\longrightarrow \infty }b(g(1/x))=b(g_{IR})  \nonumber
\end{eqnarray}

The principal result of this section will be an exact non-perturbative
formula for $b_{UV}-b_{IR}$. This will take some discussion, but
the basic
ideas are simple and we will list them here before we begin the
derivation:

\begin{enumerate}
\item  In the presence of an external source $B_{\mu }$ for the
current $J_{\mu }$, the trace anomaly of the theory includes
the familiar internal
anomaly plus an external anomaly of similar form,
\begin{equation}
\Theta =-\frac{3N_c-N_f(1-\gamma )}{32\pi^2}~
(F_{\mu\nu}^a)^2+
\frac{1}{4}q(B_{\mu \nu })^2  \label{eq:2.9}
\end{equation}
where $F_{\mu \nu }^{a}$ and
$B_{\mu \nu }=\partial _{\mu }B_{\nu}-\partial_{\nu }B_{\mu }$
are internal and external field strengths,
respectively. The general form of  (\ref{eq:2.9}) follows
uniquely from gauge invariance, locality dimensional analysis,
and parity \cite{adler}.
As discussed in refs. \cite{shiva,kogan}, the coefficient of
$(F_{\mu \nu}^a)^2$
is the numerator of the $\beta$-function (\ref{eq:2.5}).
The coefficient of $(B_{\mu \nu})^{2}$ is the
subject of our investigation. Following the discussion of
\cite{ccfis}, that
we review below, $q$ can be identified
\footnote{As discussed in Section 3, $q$ can be actually identified
with the limit as $e\to 0$ of the $U(1)$ $\beta$-function
$\beta_e (g,e)$ of an extended $SU(N_f)\times U(1)$
gauge theory with $U(1)$ coupling constant $e$,
and this provides an alternative method to derive
 the flow.
We do not use this observation here, because only the
present method can be easily extended to the gravitational
central charges
as discussed in Section 4.}
with the coefficient of a
local term
that appears in the explicit scale derivative of the regulated
correlator (\ref{jj}).
This coefficient is a scale dependent function
$\widetilde{b}(g(\mu ))$ that can be shown to have the
same $UV$ and $IR$
limits as $b(g(\mu ))$.

\item  The introduction of external sources can be done
supersymmetrically
by embedding $B_{\mu }(x)$ in an external gauge superfield
$B(x,\theta ,\bar{%
\theta})$ coupled to the superspace flavor current. The usual relation
between the trace and $\partial _{\mu }R_{\mu }$ anomalies implies that
\begin{equation}
\partial _{\mu }R_{\mu }=\frac{3N_{c}-N_{f}(1-\gamma )}{48\pi
^{2}}~F_{\mu
\nu }^{a}\widetilde{F}_{\mu \nu }^{a}-\frac{1}{6}qB_{\mu \nu
}\widetilde{B}%
_{\mu \nu }  \label{eq:2.10}
\end{equation}
In this way the flow of the central function $b(g(\mu ))$ is
related to the
flow of the anomalous correlator $\langle R_{\mu
}(x)J_{\nu }(y)J_{\rho }(z)\rangle = \frac{\partial}{\partial B_{\nu}(y)}
\frac{\partial}{\partial
B_{\rho}(z)}
\langle R_{\mu}(x) \rangle$.

\item  The final ingredient is 't Hooft anomaly matching \cite{hooft} for
the internal anomaly-free current $S_{\mu }$ in (\ref{eq:2.4}).  Since
$S_{\mu}$ is quantum-conserved in the absence of sources,
 its external anomaly is scale-independent.
This implies
that the particular combination of the external $\partial R$ and
$\partial K$
anomalies in (\ref{eq:2.4}) is independent of scale, and a useful
non-perturbative expression for the flow of $\widetilde{b}(g(\mu
))$ emerges
from this observation.
\end{enumerate}

To begin discussion of the first, and probably least familiar,
ingredient,
we consider the structure of the correlator (\ref{jj}) to any finite
order of
perturbation theory and construct a line of argument from which we
extract an
all-order result. We consider the computation of (\ref{jj}) in two
stages.
In the first stage we work at separated points and regulate all
sub-divergences at the scale $\mu $. To any finite order,
$b(g(1/x))$ can be
expressed in the form
\begin{equation}
b(g(1/x))=\sum_{n\geq 0}b_{n}(g(\mu ))t^{n},\quad ~~~~~~~~~~t=\ln
(x\mu ),
\label{eq:2.11}
\end{equation}
where $b_{n}(g)$ is a polynomial in $g$. Evaluating (\ref{eq:2.11}) at $%
x=1/\mu $, we see that $b(g)=b_{0}(g)$. The Callan-Symanzik
equations imply
\begin{equation}
\beta (g)b_{n}^{\prime }(g)+(n+1)b_{n+1}(g)=0  \label{eq:2.11.5}
\end{equation}
so that all $b_{n}(g)$ for $n\geq 1$ are proportional to $\beta
(g)$ and can
be expressed in terms of $b_{0}(g)$, $\beta (g)$ and derivatives.

When (\ref{eq:2.11}) is inserted in (\ref{jj}) one finds
$t^{n}/x^{4}$ terms
which are too singular to have a finite Fourier transform. To
correct this
we regulate this overall divergence at $x=0$ using the generalized
differential identity \cite{ccfis}
\begin{equation}
\frac{(\ln x\mu )^{n}}{x^{4}}= -\frac{n!}{2^{n+1}}{\,\lower0.9pt%
\vbox{\hrule \hbox{\vrule height
0.2 cm \hskip
0.2 cm \vrule height 0.2 cm}\hrule}\,} \sum_{k=0}^{n}\frac{
2^{k}t^{k+1}}{%
(k+1)\hbox{ ! }x^{2}}-a_{n}\delta (x),  \label{eq:2.12}
\end{equation}
which is an application of the method of differential
renormalization \cite
{diffren}. Any other method in which the overall divergence can be
separated
from sub-divergences would also work. For example, dimensional
regularization in $%
x$-space might be used. The right hand side of (\ref{eq:2.12})
contains the
Lorentz invariant solution of the differential equation ${\,\lower0.9pt%
\vbox{\hrule \hbox{\vrule height 0.2 cm \hskip 0.2
cm \vrule
height 0.2 cm}\hrule}\,} f(x)=t^{n}/x^{4}$ which is unique up to the
additive numerical constant $a_{n}$. Combining (\ref{eq:2.11}), (\ref
{eq:2.12}) and (\ref{jj}), we find the fully regulated correlator
\begin{equation}
\frac{b(g(1/x))}{x^{4}}=-\sum_{n}b_{n}(g(\mu ))\left[ \frac{n\hbox{ ! }}{
2^{n+1}}{\,\lower0.9pt%
\vbox{\hrule \hbox{\vrule height 0.2 cm
\hskip 0.2 cm
\vrule height 0.2 cm}\hrule}\,} \sum_{k=0}^{n}\frac{2^{k}t^{k+1}}{(k+1)%
\hbox{ !
}}{\frac{1}{ x^{2}}}+a_{n}\delta (x)\right] .  \label{eq:2.13}
\end{equation}

The explicit scale derivative of (\ref{jj}) gives \cite{PS} the
correlator $\langle J_{\mu}(x) J_{\nu}(0) \int d^4z \Theta(z)\rangle $
where $\Theta $ is the trace anomaly (\ref{eq:2.9}). We
compute
this scale derivative  as $\mu \,\partial /\partial \mu $ acting on
(\ref{eq:2.13}). The
result is expressed as the sum of the local contribution of the
$k=0$ term
plus a non-local term proportional to $\beta (g)$ because of
(\ref{eq:2.11.5})
\begin{equation}
\mu \frac{\partial }{\partial \mu }\frac{b(g(1/x))}{x^{4}}=2\pi^2
\widetilde{%
b}(g(\mu ))\delta (x)+\beta (g(\mu )){\,\lower0.9pt%
\vbox{\hrule
\hbox{\vrule height 0.2 cm \hskip 0.2 cm \vrule height 0.2 cm}\hrule}\,}
\frac{F(x)}{x^{2}}~,  \label{eq:2.15}
\end{equation}
where
\begin{equation}  \label{eq:2.16}
\widetilde{b} (g( \mu)) = \sum_n b_n (g(\mu))
\frac{n \hbox{ !}}{2^n}~.
\end{equation}
$F(x)$ contains the sum of all $k\geq 1$ terms in the scale
derivative, and
the $a_n$ have dropped out because they have vanishing scale
derivative. The
decomposition between local and nonlocal contributions in the above
expression is not universal. For example, one can add an arbitrary $x$
-independent function $A(g)$ to $F(x)$ and redefine
$\widetilde b(g)$ as
$\widetilde b(g)-\beta(g)A(g)$ in the local part.
However, such
contributions
are proportional to $\beta(g(\mu))$ and vanish at  the fixed points, so
the total
flow of the central function $b(g(1/x))$ can be computed from the flow of
the function $\widetilde b$ as defined in (\ref{eq:2.16}).

It is easy to derive, using (\ref{eq:2.11.5}), a differential
equation for $%
\widetilde{b}(g)$, namely
\begin{equation}
\beta (g)\frac{\partial \widetilde{b}(g)}{\partial g}+2\widetilde{b}
(g)=2b(g),  \label{eq:2.17}
\end{equation}
which is solved by
\begin{equation}
\tilde{b}(g(\mu ))=\frac{1}{\mu^2}\int_0^{\mu^2}
{\rm d}{\mu^{\prime}}^2 b(g(\mu^{\prime})).
\end{equation}
These formulas have been derived within perturbation theory, but we shall
regard them as non-perturbative results. Either formula shows that the
functions $b(g(\mu))$ and $\widetilde{b}(g(\mu))$ coincide at
fixed points
of the $RG$ flow, as well as to the second loop order in
perturbation theory
around the free fixed point \footnote{To two-loop order there are no
$\ln x \mu$ terms in (\ref{eq:2.11}) so that $\tilde{b}(g) = b(g)$.}.

Finally, one can identify $\tilde{b}(g(\mu))$
with the coefficient $q$
appearing in eq. (\ref{eq:2.9}).
To show this
we write the generating functional for the current correlation
functions as the schematic path integral
\begin{equation}
e^{-\Gamma [B_{\mu}]}=\int [d\Phi]
e^{-S[\Phi]+i\int d^4 x J_{\mu}(x) B_{\mu}(x)}.
\end{equation}
The source couples just like an abelian gauge field
without kinetic term.  The scale derivative $\mu \frac{\partial}
{\partial \mu}$ corresponds to the insertion of
$\int d^4 z \Theta (z)$ inside the path integral, so that
\begin{equation}
\label{WI}
\mu \frac{\partial}{\partial\mu} e^{-\Gamma}=
\int [d\Phi]
e^{-S[\Phi]+i\int d^4 x J_{\mu}(x) B_{\mu}(x)}
\int d^4 z  \left[-\frac{3N_c-N_f(1-\gamma)}{32\pi^2}
{}~(F^a_{\mu\nu})^2+
\frac{1}{4} q (B_{\mu\nu})^2\right]
\end{equation}
with similar internal and external contributions.
{}From  (\ref{WI}) we see that the scale derivative of the current
correlator
satisfies
\begin{eqnarray}
\label{WwI}
\mu \frac{\partial}{\partial\mu} &\langle&
J_{\mu} (x) J_{\nu} (0)
\rangle = \langle J_{\mu} (x) J_{\nu} (0) \int d^4z \Theta  (z)
\rangle \\
&=&q(\partial _{\mu}\partial _{\nu}-
\Box \delta_{\mu \nu })\delta^4(x)
-\frac{3N_c-N_f(1-\gamma)}{32\pi^2} ~ \langle J_{\mu}
(x)J_{\nu}(0)\int d^4z (F_{\mu\nu}^a)^2
\rangle.
 \nonumber
\end{eqnarray}
It has the same form as (\ref{eq:2.15}), namely the sum of a
local term
plus a non-local term proportional to $\beta (g(\mu))$, so  that we can
identify
the coefficients of the local terms, again up to contributions
$O(\beta (g(\mu ))$ that vanish at fixed points.
(Note that the last correlator in
(\ref{WwI}) is $O(g^4(\mu ))$ and vanishes in the ultraviolet.)
We can thus write
\begin{equation}
q=\frac{1}{8\pi^2}\,\widetilde{b}(g(\mu )),  \label{eq:2.24}
\end{equation}
with the understanding that possible $O(\beta (g(\mu)))$
corrections, that
are irrelevant for the total flow of $\widetilde{b}(g(\mu))$, are
omitted.

The next step is to use the $\partial_{\mu} R^{\mu}$ anomaly
(\ref{eq:2.10})
which is the supersymmetric partner of the trace anomaly
(\ref{eq:2.9}) to
compute the flow of $\widetilde{b}(g(\mu))$.
We need the anomalous
correlation functions
$\langle S_{\mu}(x) J_{\nu} (y) J_{\rho}(z)\rangle $,
$\langle R_{\mu}(x) J_{\nu} (y) J_{\rho} (z)\rangle $ and
$\langle K_{\mu}(x) J_{\nu} (y) J_{\rho} (z)\rangle $
and their flows.
The anomalous divergence of
such current correlators is one-loop exact only if all currents have no
internal anomalies. Otherwise the so-called ``rescattering
graphs,'' which
contain an internal triangle in which one of the currents
communicates to a
pair of gluons, are responsible for higher order non-local
corrections \cite
{andrei}. Therefore 't Hooft anomaly matching holds for the correlator
$\langle SJJ\rangle $ where $S_{\mu }$ is the anomaly-free current
of (\ref
{eq:2.4}), but not for $\langle RJJ\rangle $, and in general not for
$\langle KJJ\rangle $. The anomalous Ward identities of these
correlators can
be written as the following equations for matrix elements of
$S_{\mu}$, $R_{\mu}$ and $K_{\mu}$ in the presence of the
current source $B_{\mu}$
\begin{eqnarray}
\langle \partial_{\mu}S_{\mu} \rangle &\equiv &\frac{1}{48\pi^2}
s_0 B_{\mu\nu} \widetilde{B}_{\mu\nu},  \label{eq:2.26} \\[2ex]
\langle \partial_{\mu}R_{\mu} \rangle
&=&-\frac{1}{48\pi^2}\widetilde{b}(g(\mu))
B_{\mu\nu}\widetilde{B}_{\mu\nu}+\cdots ,  \nonumber
\\[2ex]
\langle \partial_{\mu}K_{\mu} \rangle
&=&-\frac{1}{16\pi^2}\tilde{k}(g(\mu))B_{\mu\nu}
\widetilde{B}_{\mu\nu}+ \cdots .  \nonumber
\end{eqnarray}
Here $s_0$ is independent of scale, while the $\partial_{\mu}K_{\mu}$
anomaly coefficient is defined as the scale-dependent function
$k(g(\mu))$.
In the $\langle\partial_{\mu}R_{\mu}\rangle$
equation "$+ ...$" indicates
the non-local contribution of the internal anomaly term of
(\ref{eq:2.10}).
There is a similar non-local contribution to
$\langle\partial_{\mu}K_{\mu}\rangle $, which cancels that of
$\langle\partial_{\mu}R_{\mu}\rangle $ in the linear combination
(\ref{eq:2.4}) that gives
$\langle\partial_{\mu }S_{\mu}\rangle .$
These contributions are irrelevant
for our analysis, since the possible local terms they contain are
$O(\beta(g(\mu)))$.

{}From (\ref{eq:2.4}) we see that the local terms in
(\ref{eq:2.26}) satisfy
\begin{equation}
\widetilde{b}(g)+\left( 1-\frac{3N_c}{N_f}-\gamma (g)\right)
\tilde{k}(g)=-s_0
{}.
\label{eq:2.27}
\end{equation}
In applications to asymptotically-free electric supersymmetric QCD, the
coupling $g(\mu )$ vanishes in the ultraviolet, and so does $\gamma (g)$;
thus the ultraviolet contributions $\tilde{b}_{UV}$ and $\tilde{k}_{UV}$
can be easily
obtained from 1-loop contributions to the relevant 3-point correlators.
These are normalized so that each quark or anti-quark field
contributes $1/N_c$ to the quantity
$\tilde{k}_{UV}$, when $J$ is the baryon current
of (\ref {eq:2.1}).
If we equate the values of the left side of (\ref{eq:2.27}) at
scale $\mu$ and in the $UV$ limit, we find
\begin{equation}
\widetilde{b}(g)=b_{UV}+\gamma (g)\tilde{k}_{UV}- \left( 1-
\frac{3N_c}{N_f}-\gamma
(g)\right) \left[ \tilde{k}(g)-\tilde{k}_{UV}\right] .  \label{eq:2.28}
\end{equation}

This result can be checked against the explicit 2-loop calculation
\cite{new}
of the flavor central function $b(g)$ and it agrees. We do not
discuss this
here, because we are more interested in non-perturbative application to
electric SUSY QCD in the conformal window,
$3N_c/2<N_f<3N_c$, where
there is evidence that the theory flows to a non-trivial fixed
point $g_{*}$ \cite{emduality}.
Although $g_{*}$ can only be calculated at the weakly
coupled end of the conformal window (where
$N_{f}=3N_{c}(1-\epsilon)$) we
know that the $\beta$-function (\ref{eq:2.5}) vanishes, and this is
enough to
give the exact infrared limit of the anomalous dimension
\begin{equation}
\gamma _{IR}=\left( 1-\frac{3N_{c}}{N_{f}}\right) .  \label{eq:2.29}
\end{equation}
Then (\ref{eq:2.28}) becomes
\begin{equation}
b_{IR}-b_{UV}= \gamma _{IR}\tilde{k}_{UV},  \label{eq:2.30}
\end{equation}
which is our first non-perturbative result for the flow of a flavor
central
charge. Since gauginos do not contribute to flavor current
correlators and
the quark and anti-quark contributions to the combination $R+K/3$
cancel, it
follows that $b_{UV}=-\tilde{k}_{UV}$ for correlators of all of the
flavor
currents
in (\ref{eq:2.1}). For the baryon current, $b_{UV}=2N_f/N_c$
and we have
the total flow
\begin{equation}
b_{IR}-b_{UV}=6\left( 1-\frac{N_f}{3N_c}\right) ,  \label{eq:2.31}
\end{equation}
which is positive in the entire conformal window,
contrary to $c$-theorem
intuition.

We now turn our attention to magnetic supersymmetric QCD with gauge
group $SU(N_c^{\prime})$, with
$N_c^{\prime }=N_f-N_c$.
The matter content
consists of $N_f$ flavors of magnetic quark and anti-quark
superfields
$q_i^{\alpha}(x,\theta)$,
$\widetilde{q}_{\alpha}^i(x,\theta)$ plus the
gauge neutral $N_f\times N_f$ meson superfield $M_j^i$.
There are
conventional gauge interactions and a cubic superpotential
$W=fqM\widetilde{q}$.
The dual theory has the same flavor group $SU(N_f)_q\times
SU(N_f)_{\widetilde{q}}\times U(1)_B$, but for simplicity we shall
consider only the baryon current \footnote{For simplicity we do
not write the scalar contributions to the
currents in
eqs.(\ref{eq:m.1}--\ref{eq:m.3}).}
\begin{equation}
J_{\mu }=\frac{1}{N_c^{\prime}}\frac{1}{2}\left(
\bar{\psi}_q\gamma_{\mu}\gamma_5\psi_q-
\bar{\psi}_{\tilde{q}}\gamma_{\mu}
\gamma_5\psi _{\tilde{q}}\right) .  \label{eq:m.1}
\end{equation}
 We also need the separate Konishi
currents of the quarks and the mesons
\begin{eqnarray}
K_{\mu}^{(q)} &=&
\frac{1}{2}\left( \bar{\psi}_{q}\gamma _{\mu }\gamma_{5}
\psi _{q}+\bar{\psi}_{\tilde{q}}\gamma _{\mu }
\gamma _{5}\psi_{\tilde{q}}\right) ,  \label{eq:m.2} \\[2ex]
K_{\mu }^{(M)} &=& \frac{1}{2}{\rm Tr}
\bar{\psi}_{M}\gamma _{\mu }\gamma_{5}\psi_{M},  \nonumber
\end{eqnarray}
while the fermion content of the $R$ current is
\begin{equation}
R_{\mu }= \frac{1}{2}\bar{\lambda}^a\gamma_{\mu}
\gamma_5\lambda^a-
\frac{1}{3}(K_{\mu}^{(q)}+K_{\mu}^{(M)}).  \label{eq:m.3}
\end{equation}
The divergences $\partial_{\mu} K_{\mu}^{(q)}$ and
$\partial_{\mu}K_{\mu}^{(M)}$ have classical contributions,
since the
superpotential is not
invariant under the relevant $U(1)$ transformations,
and $\partial_{\mu}K_{\mu}^{(q)}$ also has an anomaly.
The internal $R$-current anomaly also
involves $F\tilde{F}$ and there is a superpotential contribution \cite
{kogan,strassler}.
It is significant for our analysis that there is
a unique
combination \cite{kogan} of these currents which is  classically
conserved
and anomaly-free, namely
\begin{equation}
S_{\mu }=R_{\mu}+\frac{1}{3}\left(
1-\frac{3N_c^{\prime}}{N_f}-\gamma_{q}\right)
\left( K_{\mu}^{(q)}-2K_{\mu}^{(M)}\right)
-\frac{1}{3}\left(
2\gamma _{q}+\gamma _{M}\right) K_{\mu }^{(M)}  \label{eq:m.4}
\end{equation}
The combination $K_{\mu}^{(q)}-2K_{\mu}^{(M)}$ is conserved
classically.
Its coefficient in (\ref{eq:m.4}) is again the numerator of the
NSVZ gauge
$\beta$-function (\ref{eq:2.5}),
and the coefficient of $K_{\mu }^{(M)}$ is
essentially the Yukawa  $\beta$-function
$\beta_f=f(2\gamma_q+\gamma_M)$.

The previous analysis must be generalized to include a
superpotential. There
are only trivial changes in the derivation of the relation between the
external trace anomaly $\widetilde{b}$ and the central function
$b(g(1/x)$,
$f(1/x))$, and the result that they coincide at
fixed points is unchanged.
The analysis of the flow of flavor central functions is also easily
repeated. The third equation of (\ref{eq:2.26}) is replaced by the pair
\begin{eqnarray}
\langle\partial_{\mu }K_{\mu }^{(q)} \rangle &=&
-\frac{1}{16\pi^2}\tilde{k}^{(q)}\left( g(\mu),f(\mu )\right)
B_{\mu \nu }\widetilde{B}_{\mu\nu} +...
\label{eq:m.6} \\[2ex]
\langle\partial _{\mu }K_{\mu}^{(M)}\rangle &=&
-\frac{1}{16\pi^2}
\tilde{k}^{(M)}\left( g(\mu),f(\mu )\right)
B_{\mu\nu}\widetilde{B}_{\mu\nu } +...
\nonumber
\end{eqnarray}
and one finds
\begin{eqnarray}  \label{HH}
\widetilde{b} &=&b_{UV}+\gamma _{q}\tilde{k}_{UV}^{(q)}
+\gamma _{M}\tilde{k}_{UV}^{(M)}
\label{eq:m.7} \\[2ex]
&&-\left( 1-\frac{3N_c^{\prime}}{N_f}-\gamma_q\right)
\left(\tilde{k}^{(q)}-2\tilde{k}^{(M)}-\tilde{k}_{UV}^{(q)}
+2\tilde{k}_{UV}^{(M)}\right)  \nonumber \\[2ex]
&&+\left( 2\gamma _{q}+\gamma _{M}\right) \left(
\tilde{k}^{(M)}-\tilde{k}_{UV}^{(M)}\right)
\nonumber
\end{eqnarray}
in which the subscript $UV$ denotes a quantity evaluated in the
ultraviolet
from 1-loop graphs, and other quantities are evaluated at scale
$\mu $. One
can also check, as in the electric case, that this formula agrees
with the
two-loop calculation.

We now go to the $IR$ fixed point where the coefficients of the last two
terms of (\ref{eq:m.7}) vanish, since they are related to $\beta
_{g}$ and $%
\beta _{f}$, respectively. This gives the $IR$ values of the anomalous
dimensions
\begin{equation}
\gamma _{q}^{IR}=-\frac{1}{2}\gamma _{M}^{IR}=\left(
1-\frac{3N_{c}^{\prime
} }{N_{f}}\right)  \label{eq:m.8}
\end{equation}
and we obtain from (\ref{eq:m.7})
\begin{equation}
b_{IR}-b_{UV}=\left( \gamma _{q}^{IR}\tilde{k}_{UV}^{(q)}+\gamma
_{M}^{IR}\tilde{k}_{UV}^{(M)}\right) ,  \label{eq:m.9}
\end{equation}
which is a non-perturbative formula for the flow of
flavor charges in the
magnetic theory. For the baryon current, $b_{UV}=-
\tilde{k}_{UV}^{(q)}=2N_{c}/N_{c}^{\prime }$ and
$\tilde{k}_{UV}^{(M)}=0$,
 so the flow of the baryon central charge is
\begin{equation}
b_{IR}-b_{UV}=6\left( 1-\frac{N_f}{3N_c^{\prime }}\right) =2\,
\frac{2N_f-3N_c}{N_{f}-N_c}  \nonumber  \label{eq:m.10}
\end{equation}
which is again positive throughout the conformal window.
{}From
(\ref{eq:2.31}) and (\ref{eq:m.10}) one can check the
equality of the $IR$ values
of the
central function, namely $b_{IR}=6$.
This is not a new confirmation of
duality, since it is the same 't Hooft anomaly matched in
electric-magnetic
duality \cite{emduality}.

We briefly discuss the question of constancy of the flavor central
charge on conformal fixed lines \cite{strassler}, confining our
 attention to the $N=4$ and
$N=2$ cases. The matter content of the $N=4$ theory is  three
adjoint  chiral
multiplets, while for $N=2$, with $G=SU(N_c)$, $N_f=2N_c$
fundamental hypermultiplets are
required to produce a fixed line. In each case there is a
particular cubic
superpotential, and the condition for the fixed line is
a linear relation
between $g$ and the Yukawa coupling $f$. It is straightforward to
generalize
the previous derivations and show that formulas of the type
(\ref{eq:2.30})
and (\ref{eq:m.9}) hold for the $N=4$ and $N=2$ cases,
respectively, with
$UV$-subscripted quantities calculated from the
relevant 1-loop graphs.
To
establish marginal constancy at the non-perturbative level,
we need only
observe \cite{strassler} that anomalous dimensions vanish on the
fixed line,
so we have $b=$constant.
At first thought one might expect that the
value of
the central charge depends on the single coupling, say $g$, that
remains on
the fixed line; however we see that $b$ is
determined by the free field
content of the theory.

An even simpler argument is to note \cite{kogan}
that $S_{\mu }$ and $R_{\mu}$ coincide at infrared fixed
points or fixed lines. But $S_{\mu }$ is
anomaly-free, so we have
$\langle \partial _{\mu }S_{\mu }J_{\nu }J_{\rho}\rangle _{UV}=
\langle \partial _{\mu }S_{\mu }J_{\nu }J_{\rho }\rangle_{IR}=
\langle \partial _{\mu }R_{\mu }J_{\nu }J_{\rho }\rangle _{IR}$.
Hence, the central charge $b$ on an infrared fixed line is equal to the
anomaly coefficient $-s_0$ and thus depends only on the
free-field content
of the theory and not on any coupling constant.

\section{An alternative approach}
\setcounter{equation}{0}

In this section
we compute the flavor central charges using a somewhat
different approach.
This method uses the correlation function
$b(g(1/x))$ of (\ref{jj}) in a more direct way, and does not
require an explicit regularization of the singularity at $x=0$.
This simplifies the analysis.
The method is quite general, but
we work with the baryon number current of the electric SUSY QCD,
given in (\ref{eq:2.1}) for simplicity.

Let us consider the correlator (\ref{jj}) again and compute the
scale derivative
$\mu\partial/\partial \mu \langle J_{\mu}(x)
J_{\nu}(0)\rangle$ at fixed bare coupling $g_0=g(\mu)$,
which gives by definition the correlator
$\langle J_{\mu}(x)J_{\nu }(0)\int d^4 z ~ \Theta (z)\rangle$.
In the presence of an external field $B_\mu$ coupled
to $J_\mu$ and in quadratic approximation
the matrix element
$<\int d^4 z~\Theta (z)>$ has the form
\begin{eqnarray}
\label{Aone}
<\int d^4 z~\Theta (z)> &=& \int \frac{d^4 k}{(2\pi)^4}~
f(g(k))\cdot\frac{1}{4} B^{\mu
\nu }(k)B_{\mu\nu}(-k)=  \label{der} \\
&=&\int \frac{d^4 k}{(2\pi)^4}~\frac{1}{4}
B^{\mu \nu }(k) B_{\mu \nu }(-k) \left( f_0-
\frac{1}{16\pi^4}
\int d^4 x {\rm e}^{ikx}\frac{1}{x^4}\mu
\frac{\partial }{\partial \mu } b(g(1/x))\right) =
\nonumber \\
&=&\int \frac{d^4 k}{(2\pi)^4}~\frac{1}{4}
B^{\mu \nu }(k)B_{\mu \nu }(-k) \left( f_0-
\frac{1}{8\pi^4}\int d^4 x {\rm e
}^{ikx}\frac{1}{x^4}\frac{\partial }{\partial \log x^2}
b(g(1/x))\right) .
\nonumber
\end{eqnarray}
The above integral is convergent at $x=0$ because of asymptotic
freedom and the RG equations.
In eq. (\ref{der}) we explicitly
introduced a constant $f_{0}$ associated with the regularization
of the correlator (\ref{jj}) at $x=0$.
This contact term is irrelevant for the present analysis.
However it can be easily determined.
Indeed, in the UV limit, i.e. $|k|\to \infty$, the
$x$-integral in eq. (\ref{der}) vanishes, and hence $f_0=f_{UV}=
f(g(k))\mid_{|k|\to \infty}.$
In particular in the case of the
baryonic current $f_0=N_f/ 4\pi^2 N_c.$

The IR limit corresponds to $|k|\to 0,$ i.e.
$f_{IR}=f(g(k))\mid_{|k|\to 0}$.
Thus from (\ref{der}) one gets
\begin{equation}
f_{UV}-f_{IR}=-\frac{1}{8\pi^2}\int_{0}^{\infty }d\log x^2~
\frac{\partial
}{\partial
\log x^2}b(g(1/x))= \frac{1}{8\pi^2}(b_{UV}-b_{IR}).  \label{final}
\end{equation}
We turn now to the computation of the function $f(g(k))$.
The simplest way is to observe that the
$B_\mu J_\mu$ coupling is essentially that of SQED,
with external
``photon'' field $B_\mu$ and coupling constant $e$.
In SQED the conformal anomaly can be computed as a scale
derivative $\mu\partial/\partial\mu$ (at fixed values of $g(\mu),
{}~e(\mu)$)
of the effective action for the external field $B_{\mu}$
\begin{equation}
\Gamma[B]=\int \frac{d^4 k}{(2\pi)^4} ~\frac{1}{4e^2(k)}
B_{\mu\nu}(k)B^{\mu\nu}(-k),
\end{equation}
where $e(k)$ is the effective running coupling constant.
We have
\begin{eqnarray}
\label{TQED}
<\int d^4 z~\Theta (z)>=\mu\frac{\partial}{\partial\mu}
\Gamma[B]&=& \frac{1}{N_c^2}\int \frac{d^4 k}{(2\pi)^4}
B(k)^{\mu \nu }{B}_{\mu \nu}(-k)\cdot
\frac{\beta_e(e(k), g(k))}{2e^3(k)}\cr &=& \frac{1}{N_c^2}
\int \frac{d^4 k}{(2\pi)^4}
B(k)^{\mu \nu }{B}_{\mu \nu }(-k)\cdot \frac{N_f N_c}{16\pi^2}~[1-\gamma
(g(k))].
\end{eqnarray}
This expression follows from the general result for the N=1
supersymmetric QED  $\beta$-function
$\beta_e=n \,(1-\gamma)e^3/8\pi^2$ given in
ref. \cite{nsvz},
$\gamma $ being the anomalous dimension of the matter
chiral superfields and $n$ standing for the number of flavors.
In the present context we have $n=N_{c}N_{f}$, and $e\to 0$
because the QED gauge field $B_{\mu}$ is external.
Therefore,
\begin{equation}
f(g(k))=\frac{N_f}{4\pi^2 N_c}~[1-\gamma
(g(k))]. \label{ff}
\end{equation}
gives an exact formula for $f(g(k))$ as a linear
function of the anomalous dimension. (The relation (\ref{ff}) is not scheme
independent, but rather holds in the scheme of \cite{kogan}.)
The inverse Fourier
transform  of (\ref{ff}) then gives
$\mu \frac{\partial}{\partial \mu} \left(b(g(1/x)) /x^4\right)$,
and $b(g(1/x))$ can be easily computed.

Since $\gamma(g(k))\mid_{k\to \infty }=0$ and
$\gamma (g(k))\mid _{k\to 0}=1-3N_c/N_f$,
from eq. (\ref{ff}) immediately follows that
\begin{equation}
b_{IR}-b_{UV}=8\pi^2(f_{IR}-f_{UV})=2
\cdot\frac{N_f}{N_c}\cdot \left(
\frac{3N_c}{N_f}-1\right) .
\end{equation}
We therefore recover the result of Section 2.
Note that we did
not use the current $S_{\mu }$ of (\ref{eq:2.4}).

It is worth observing that the SQED  $\beta$-function
is also implicitly present in the treatment of Section 2.
To show this we consider
a theory with gauge group extended to
$SU(N_c)\times U(1)$ with $U(1)$
coupling constant $e$.
This theory has the operator trace anomaly
\begin{equation}
\Theta= -\frac{3N_c-N_f(1-\gamma)}{32\pi^2}
\left(F^a_{\mu\nu}\right)^2+
\frac{N_c N_f (1-\gamma)}{16\pi^2 N_c^2}
\left(B_{\mu\nu}\right)^2, \label{twice}
\end{equation}
where the coefficient of $\left(B_{\mu\nu}\right)^2$
is $\beta_e/2e^3N_c^2$ and $\gamma=\gamma(e(\mu),g(\mu))$.
In the limit $e\to 0$ we can compare
(\ref{ff}) and (\ref{twice}) with
(\ref{eq:2.9}) and identify
\begin{equation}
q=\lim_{e\to 0} \frac{2\beta_e}{e^3N_c^2}=
\frac{N_c N_f (1-\gamma (0,g(\mu))}{4\pi^2 N_c^2}
=f(g(\mu ))
\end{equation}

It is instructive to compare  (\ref{der}) and (2.21).
Consider first  (2.21).
The correlator $\langle J_{\mu}
J_{\nu}\int \Theta\rangle$ is renormalization group invariant.
In momentum space
the r.h.s of  (2.21) may depend only on the running
coupling $g(k)$,  but
not on $g(\mu)$.
Therefore, the $\mu$-dependence of $q$ is
compensated by that of the correlator $\langle J_{\mu} J_{\nu}\int
(F^a_{\mu\nu})^2\rangle$ which has a perturbative expansion
starting with $g^4(\mu)\log k^2/\mu^2$.
On the other hand the
form of   (\ref{der}) is explicitly renormalization group
invariant.
The local part $q$ of  (2.21) was identified above as
$q(g(\mu))=f(g(\mu)).$
The non-local part of
(\ref{der}) corresponding to the contribution of $\langle
J_{\mu} J_{\nu}\int (F^a_{\mu\nu})^2\rangle$ in (2.21), is then
\begin{equation}
\label{eq:2}
f(g(k))-f(g(\mu))\sim
[3N_c-N_f(1-\gamma(g(\mu)))]\cdot
[g^4(\mu)\log k^2/\mu^2 +\ldots]\,.
\end{equation}
This is proportional to $\beta(g(\mu))$ because of the RG
equation
\begin{equation}
  \label{eq:3}
  \mu \, \frac{\partial}{\partial \mu} b(g(1/x)) = \beta (g(\mu))
  \, \frac{\partial}{\partial g(\mu)} b (g(1/x))
\end{equation}
which can be used in (\ref{der}), and perturbation theory gives
the leading power $g^4(\mu)$. In this way we establish a
correspondence between the approaches of Section 2 and the present
section.

Another  method of calculation  is to consider first the anomalous
superspace operator equations in the absence of the external
$B_{\mu }$ field (see, for example \cite{nsvz,noialtri})
\begin{equation}
\bar{D}^{\dot{\alpha}}J_{\alpha \dot{\alpha}}-
\frac{N_f}{48\pi^2}
\left( 1-\frac{3N_c}{N_f}-\gamma (g)\right)
D_{\alpha}W^2=0,{}~~~~
\bar{D}^2K-\frac{N_f}{2\pi^2}W^2=0,  \label{SSaneq}
\end{equation}
where $J_{\alpha \dot{\alpha}}$ is the supercurrent, $K$ is the
Konishi operator.
By combining eqs. (\ref{SSaneq}) one can get
\[
\bar{D}^{\dot{\alpha}}J_{\alpha \dot{\alpha}}-
\frac{1}{24}\left( 1-
\frac{3N_c}{N_f}-\gamma (g)\right) D_{\alpha }\bar{D}^{2}K=0,
\]
which is the superspace version of (\ref{eq:2.4}).  In the
presence of the external field $B_{\mu }$ the operator
$\bar{D}^{\dot{\alpha}}J_{\alpha \dot{\alpha}}-
\frac{1}{24}\left(1-\frac{3N_c}{N_f}- \gamma
(g)\right)D_{\alpha}\bar{D}^2 K$ does not vanish but is
proportional to the unit operator times a local functional of
the external field.
Hence, for the component matrix element
containing the stress tensor we have ($ \Theta\propto
[D^{\alpha}, \bar{D}^{\dot{\alpha}}] J_{\alpha
\dot{\alpha}}|_{\theta =0}$, $L_{matter}\propto \{D^2,
\bar{D}^2]K\}_{\theta =0}$ )
\begin{equation}
<\int d^4
z~\left[ \Theta (z)-\left( 1-{\frac{3N_c}{N_f}}-\gamma
(g)\right) L_{matter}(z)\right] >=
u_0\int \frac{d^4 k}{(2\pi)^4}
\frac{1}{4} B(k)^{\mu\nu}{B}_{\mu\nu}(-k).  \label{Tnull}
\end{equation}
The constant $u_0$ does not depend on
the external field.
It is also independent of the renormalization group
scale $\mu$ since the
operator on the left hand side is renormalization group
invariant.

(It is worth noting that the above statement is a consequence of
a general result.
We refer to an operator $O$ that has vanishing
matrix elements between any Fock physical states as a {\it null}
operator.
This is described by the operator equation $O=0$.
Thus,
$\bar{D}^{\dot{\alpha}}J_{\alpha \dot{\alpha}}-
\frac{1}{24} ( 1-3N_c/N_f-\gamma ) D_{\alpha }
\bar{D}^2 K$ is a null operator.
In the presence of an external field a
null operator may not vanish.
In general the r.h.s. of the
operator equation may be a linear combination of various local
operators $O_i$ with the coefficients $y_i$ being non-trivial
local functionals of the
external field, i.e.  $O=\sum_i y_i O_i$.
In the simplest case the $O$ operator mixes only with the unit
operator,
i.e. the operator equation reads $O=y \cdot {\bf 1}$.
In such a
case the matrix element of $O$ is just a local functional $y$ of
the external field.  Note that the one-loop form of the 't~Hooft
external anomaly for a current $J_{\mu}$ is a particular
consequence of this fact.  Indeed, assuming that $J_{\mu}$ does
not have any internal anomaly its anomalous dimension is
vanishing.  Therefore the matrix element of
$\partial_{\mu}J_{\mu}$ is renormalization group invariant, i.e.
it may depend only on the running coupling $g(k)$, where $k$ is a
momentum of the external field.
On the other hand, as mentioned
above, $y$ is a local functional of the
external field, and hence it does not depend on $g(k)$.
Therefore the matrix element in
question is exactly one-loop.)

For technical reasons it is convenient
to consider a particular momentum mode $k$ and
define the reduced matrix elements
$\ll\cdots\gg $ obtained by dividing the conventional matrix
elements $\langle \cdots \rangle $ by
$(1/4)B^{\mu\nu}(k)B_{\mu\nu}(-k).$
The reduced matrix element of the
operator
$\int d^4z~[\Theta (z)-(1-3N_c/N_f-
\gamma(g))~L_{matter}(z)]$ (which is just equal to $u_0$)
does not
change under the renormalization group flow with respect to
$k^2$.
Therefore we have
\begin{eqnarray}
\Delta f=f(g(k))\mid _{|k|\to 0}- &\;&f(g(k))\mid _{|k| \to
\infty } =\ll \int d^4 z~(1-3N_c/N_f-\gamma (g(\mu )))
L_{matter}(z)\gg \mid_{|k|\to 0}-\cr
&\;&\ll \int d^4 z (1-3N_c/N_f-\gamma
(g(\mu ))) L_{matter}(z)\gg\mid _{|k|\to \infty }.  \label{done}
\end{eqnarray}
The matrix element $\ll \int d^4 z~(1-3N_c/N_f-\gamma (g))~
L_{matter}(z)\gg$ is proportional to
$1-3N_{c}/N_{f}-\gamma (g(k))$ \cite{noialtri} and
hence vanishes in the infrared, i.e. at $k^2\to 0$.
Thus we have, precisely as before,
\begin{eqnarray}
\Delta f&=&-\ll \int d^4 z~(1-3N_c/N_f-\gamma (g))~
L_{matter}(z)\gg \mid_{|k|\to \infty }=  \label{dtwo}
\nonumber\\
&=&\frac{1}{4\pi^2} \left( \frac{3N_c}{N_f}
-1\right) \frac{N_f}{ N_c},
\end{eqnarray}
where the result follows from the identification of $L_{matter}$
 with the $\theta^2 \bar{\theta}^2$ component of the Konishi
superfield, and the  (one-loop) Konishi anomaly.

The story becomes more complicated for the gravitational central
charges which are considered in section 4.
In this case one may try to analyse the matrix element $<\int d^4
z~\Theta (z)>$
of the stress tensor in the presence of an external gauge field
$V_{\mu}$ field
coupled to the $R_{\mu }$ current in order to compute the flow,
say, of the $c$ charge.
However, the coupling of the $R_{\mu }$ current to the ``QED''
vector field $V_{\mu }$ breaks supersymmetry and this ``QED''
$\beta$-function
is not simply related to the anomalous dimensions $\gamma $.
A direct way to
compute the flows of the $c$ and $a$ charges is to consider the matrix
element of the divergence of the $R_{\mu }$ current in the presence of an
external gauge field coupled to the $R_{\mu }$ current. In this way it is
easy to rephrase the considerations of Section 4 in the formalism of this
section.

\section{Gravitational central functions}

\setcounter{equation}{0}

As discussed in \cite{ccfis} the correlator of two stress tensors can be
written in the form
\begin{equation}  \label{eq:3.1}
\langle T_{\mu \nu }(x)\,T_{\rho \sigma }(0) \rangle
=-{\frac{1}{48\pi ^{4}}}
\Pi _{\mu \nu \rho \sigma } \frac{c(g(1/x))}{x^4}
+\pi _{\mu \nu }\pi
_{\rho \sigma } \frac{f(\ln x\mu ,g(1/x ))}{x^4} ,  \label{decomp}
\end{equation}
where $\pi _{\mu \nu }=
(\partial _{\mu }\partial _{\nu }-
\delta_{\mu \nu }{\,\lower0.9pt
\vbox{\hrule \hbox{\vrule height 0.2 cm \hskip 0.2
cm \vrule
height 0.2 cm}\hrule}\,} )$ and
$\Pi _{\mu \nu \rho \sigma }=2\pi_{\mu \nu
}\pi _{\rho \sigma }-3(\pi _{\mu \rho }\pi _{\nu \sigma }+
\pi _{\mu\sigma
}\pi _{\nu \rho })$ is the transverse traceless spin 2 projector.
One of the
objects of primary concern in this section is the flow of the central
function $c(g(1/x))$ that we will relate to the coefficient of the
square of
the Weyl tensor in the external gravitational trace anomaly. The second
object of interest is the coefficient $a$ of the Euler term in the trace
anomaly. We will give non-perturbative formulas for the flows of these
quantities. The internal trace anomaly is responsible for the
second tensor
structure in (\ref{eq:3.1}). It is proportional to $\beta (g(\mu ))$, and
thus vanishes at critical points.

Since $T_{\mu\nu}(x)$ and $R_{\mu}(x) $ are both components of the
supercurrent superfield $%
J_{\alpha\dot\alpha}(z)$,  $z \equiv (x, \theta , \bar{\theta})$,
there is a relation between the $\langle TT
\rangle$
and $\langle RR \rangle$ correlators. For a critical supersymmetric
theory,
they are both contained in the supercorrelator
\begin{equation}  \label{eq:4.2}
\langle J_{\alpha \dot{\alpha}}(z) J_{\beta \dot{\beta}}(0)\rangle
\sim {c}%
\,\,{\frac{ s_{\alpha\dot{\beta}} \bar{s}_{\beta \dot{\alpha}}
}{(s^2 \bar{s}%
^2)^2}}
\end{equation}
given in Section 5 of \cite{noialtri} where the notation is
explained. This
means that the central charges of the $TT$ and $RR$ OPE's are given
by the
same constant $c$ which is a fixed point value of $c(g(\mu ))$. Off
criticality there is a more complicated relation between $\langle
TT \rangle$
and $\langle RR \rangle$ that is not required for our work.

In Section 2 we introduced a source for the flavor current $%
J_{\mu}(x)$ in order to show that the coefficient of the external trace
anomaly coincides at fixed points with the central function
$b(g(1/x))$ and
 related the trace and $\partial_{\mu} R_{\mu}$ anomalies
using global
supersymmetry. In the present case sources for $T_{\mu\nu}$ and $R_{\mu}$
take us out of the realm of global supersymmetry and require
external field
supergravity where the anomaly situation is more complex, as we now
discuss.

We introduce the background metric $g_{\mu \nu }(x)$ and source
$V_{\mu }(x)$
for the $R$-current. In these background fields the trace anomaly of a
critical supersymmetric theory has the  form
\begin{equation}  \label{eq:4.2.5}
\Theta
=\frac{c}{16 \pi^2}\,(W_{\mu \nu \rho \sigma})^{2}-\frac{a}{16 \pi^2}\,
\widetilde{R}_{\mu \nu \rho \sigma}\widetilde{R}^{\mu \nu \rho \sigma}
+\frac{c}{6 \pi^2}\,V_{\mu \nu
}^{2}~.  \label{1.}
\end{equation}
Here $W_{\mu \nu \rho \sigma} $ is the Weyl tensor and
 $\widetilde{R}_{\mu \nu \rho \sigma}$ is the dual of the
curvature
tensor, the second term being the Euler density; $V_{\mu \nu}$ is the
 field strength of $V_{\mu}$. The coefficients
of the $(W_{\mu \nu \rho \sigma})^{2}$ and $(V_{\mu \nu })^{2}$ terms
 are related while that of the
Euler density is an independent constant.
In a free supersymmetric gauge theory with $N_{v}$ gauge and $N_{\chi }$
chiral multiplets, the constants are \cite{CD}
\begin{equation}
c_{UV}=\frac{1}{24}\left( 3N_{v}+N_{\chi }\right)
{}~~,{}~~~a_{UV}=\frac{1}{%
48}\left( 9N_{v}+N_{\chi }\right) ~.  \label{ca}
\end{equation}
Off criticality there are additional terms in $\Theta $ that are
proportional to $\beta (g(\mu ))$ and do not contribute to the
total flow,
and the central charges depend on the coupling, i.e. $c=c(g(\mu
))$ and $%
a=a(g(\mu ))$.

In the superspace description of these anomalies, the external metric
(actually the vierbein $e_{\mu }^{~a}$) and current source (the
supergravity
axial vector auxiliary field) are contained in a single superfield
$H^a(x,
\theta , \bar{\theta})$, and the trace anomaly and the $\partial_{\mu}
R_{\mu}$ anomaly are components of the superfield equation
\begin{equation}
\bar{D}^{\dot{\alpha}}J_{\alpha \dot{\alpha}}=D_{\alpha }J
\label{eq:4.3}
\end{equation}
where the chiral superfield $J$ (the supertrace) has the form
\begin{equation}
J = \frac{1}{24 \pi^2}(c W^2 - a \Xi_c)  \label{eq:supertrace}
\end{equation}
in terms of the {\em super}Weyl   tensor $W_{\alpha \beta \gamma}$ and the
chirally
 projected {\em super}Euler density (see Appendix A for details).

We need the $\partial _{\mu }R_{\mu }$ anomaly. Although it is fairly
straightforward to compute it as the theta component
of (\ref{eq:4.3}), and we will do this in Appendix A, we
present here an
indirect argument
that uses the general structure of (\ref{eq:supertrace}) but does not
require any superspace technology. The $\partial _{\mu }R_{\mu }$ anomaly
will have the form
$\partial _{\mu }R_{\mu }=(uc+va)R_{abcd}\widetilde{R}^{abcd}+
(wc+za)V\widetilde{V}$ with model independent coefficients $u$, $v$,
$w$ and $z$. It is thus sufficient to compute in a free
supersymmetric gauge
theory in order to find these coefficients. Because of the ratio $-1/3$
between the $R$-charges of gauginos and matter fermions, we know
that $uc+va$
can be obtained from the $\langle TTR\rangle $ triangle graph as a pure
numerical (i.e. $1/24\pi ^{2}$) multiple of $3N_{v}-N_{\chi }$,
and $wc+za$
can be obtained from the $\langle RRR\rangle $ triangle as a numerical
multiple of $27N_{v}-N_{\chi }$. Using the values for $c_{UV}$ and
$a_{UV}$
in (\ref{ca}) we find
\begin{equation}
\partial _{\mu }(\sqrt{g}R^{\mu })=\frac{c-a}{24 \pi^2}
R_{\mu \nu \rho \sigma}\widetilde{R}^{\mu \nu \rho \sigma}
+\frac{5a-3c}{9\pi^2}V_{\mu \nu}
\widetilde{V}^{\mu \nu}  \label{eq:4.4}
\end{equation}

The next step is to note that the $W^{2}$ anomaly coefficient of a
non-critical interacting theory, $\tilde{c}(g(\mu ))$, can be
related to the
central function $c(g(1/x))$ by the same argument as in Section 2,
which in
fact is exactly the argument given in \cite{ccfis}. (The Euler term of $%
\Theta $ gives no contribution to the integrated trace anomaly.) The
functions $\tilde{c}(g(\mu ))$ and $c(g(\mu ))$ coincide at fixed
points of
the flow.

We now proceed to the calculations of $c$ and $a$ in the electric
$SU(N_{c})$
SQCD, where $N_{v}=N_{c}^{2}-1$ and $N_{\chi }=2N_{c}N_{f}$.
To achieve this goal, we can apply arguments of
Section 2 to
calculate the flow of the combinations of anomaly coefficients $\tilde{c}
(g(\mu ))-{a}(g(\mu ))$ and $5{a}(g(\mu
))-3\tilde{c}(g(\mu ))$
from the three-point correlators $\langle TTR\rangle $ and $\langle
RRR\rangle $, respectively.

The anomalies of the correlators $\langle TTS\rangle $, $\langle
TTR\rangle$,
and $\langle TTK\rangle $ are summarized by the equations
\begin{eqnarray}
\langle\partial _{\mu }(\sqrt{g}S^{\mu })\rangle &=&
\frac{1}{12\pi ^{2}}s_{1}\epsilon
^{\mu \nu \rho \sigma }R_{\mu \nu \lambda \tau }R_{\rho \sigma }^{\lambda
\tau },  \label{eq:4.5}  \nonumber\\[2ex]
\langle\partial _{\mu }(\sqrt{g}R^{\mu })\rangle &=&\frac{1}{12\pi
^{2}}\left(
\tilde{c}%
(g(\mu ))-{a}(g(\mu ))\right) \epsilon ^{\mu \nu \rho \sigma
}R_{\mu
\nu \lambda \tau }R_{\rho \sigma }^{\lambda \tau },  \nonumber
\nonumber\\[2ex]
\langle\partial _{\mu }(\sqrt{g}K^{\mu }) \rangle &=&\frac{1}{12\pi
^{2}}k(g(\mu
))\epsilon ^{\mu \nu \rho \sigma }R_{\mu \nu \lambda \tau }R_{\rho \sigma
}^{\lambda \tau }, \end{eqnarray}
where we omitted the non-local $O(\beta (g(\mu)))$ terms.
In the ultraviolet limit the quantity $k(g(\mu ))$ can be
obtained
from the 1-loop triangle graph as $k\to k_{UV}=-N_{f}N_{c}/8$.
Similarly $\tilde{c}(g(\mu ))-{a}(g(\mu ))\to
c_{UV}-a_{UV}=-\frac{1}{16}(N_{c}^{2}-1-\frac{2}{3}N_{f}N_{c})$.
It is now immediate to write the
analogue of (\ref{eq:2.28}) which is
\begin{equation}
\tilde{c}(g)-a(g)=c_{UV}-a_{UV}+\frac{1}{3}\gamma
(g)k_{UV}-\frac{1}{3}\left( 1-
\frac{3N_{c}}{N_{f}}-\gamma (g)\right) (k(g)-k_{UV}).  \label{eq:4.6}
\end{equation}

To discuss the $\langle RRR\rangle $ anomaly, we observe that the
amplitude
of the triangle graph for the contribution of one Majorana spinor  with
current  $J_{\mu} = \frac{1}{2} \bar{\psi} \gamma_{\mu}\gamma_5 \psi$
 has the Bose-symmetric
anomaly
\begin{equation}
\frac{\partial }{\partial z_{\rho }}\langle J_{\mu }(x)J_{\nu }(y)
J_{\rho}(z)\rangle _{UV}=-\frac{1}{12\pi ^{2}}\epsilon _{\mu \nu
\rho \sigma }
\frac{\partial }{\partial x_{\rho }}\frac{\partial }{\partial y_{\sigma
}}\delta
(x-z)\delta (y-z)\equiv\frac{16}{9} {\cal A}_{\mu \nu }(x,y,z)
\label{eq:5.9}
\end{equation}
(see \cite{erlich} for a recent discussion of the anomaly in
$x$-space). We
then write $R_{\mu }=S_{\mu }+\frac{1}{3}(\gamma -\gamma _{IR})K_{\mu }$
where $S_{\mu }$ is the internal anomaly-free current (\ref{eq:2.4}).
The various Bose symmetric contributions to $\langle RRR\rangle $
then have
 anomalous divergences that can be
written as:
\begin{eqnarray}
&&\frac{\partial }{\partial z_{\rho }}\langle R_{\mu }R_{\nu }R_{\rho
}\rangle
=[5a(g(\mu ))-3c(g(\mu )]{\cal A}_{\mu \nu } \nonumber \\[0.03in]
&&\frac{\partial }{\partial z_{\rho }}\langle S_{\mu }S_{\nu }S_{\rho
}\rangle
=s_{3}\,{\cal A}_{\mu \nu } \nonumber \\[0.03in]
&&\frac{\partial }{\partial z_{\rho }}\left[ \langle S_{\mu }S_{\nu
}K_{\rho
}\rangle +\langle S_{\mu }K_{\nu }S_{\rho }\rangle +\langle K_{\mu
}S_{\nu
}S_{\rho }\rangle \right]  =3k_{1}(g(\mu ))\,{\cal A}_{\mu \nu }
\nonumber\\[2ex]
&&\frac{\partial }{\partial z_{\rho }}\left[ \langle S_{\mu }K_{\nu
}S_{\rho
}\rangle +\langle K_{\mu }K_{\nu }S_{\rho }\rangle +\langle K_{\mu
}S_{\nu
}K_{\rho }\rangle \right]  =3k_{2}(g(\mu )){\cal A}_{\mu \nu }  \nonumber
\\[2ex]
&&\frac{\partial }{\partial z_{\rho }}\langle K_{\mu }K_{\nu }K_{\rho
}\rangle
=k_{3}(g(\mu )){\cal A}_{\mu \nu }
\end{eqnarray}
where the tensor indices $\mu $, $\nu $, $\rho $, are associated with
coordinates $x$, $y$, $z$, respectively;
we again omitted $O(\beta )$ irrelevant nonlocal terms.
The  anomaly coefficient  $s_{3}$ is
scale independent while other anomaly coefficients depend on $g(\mu )$.


Proceeding as before and evaluating $s_{3}$ in the $UV$ limit from
triangle
graphs, we obtain
\begin{equation}
5a-3c=5a_{UV}-3c_{UV}+h,  \label{gt}
\end{equation}
where
\begin{eqnarray}
h&=&\gamma _{IR}k_{1\,UV}-\gamma _{IR}^{2}\frac{1}{3}k_{2\,UV}+\gamma
_{IR}^{3}%
\frac{1}{27}k_{3\,UV}+(\gamma -\gamma _{IR})k_{1} \nonumber\\
&&+(\gamma -\gamma
_{IR})^{2}%
\frac{1}{3}k_{2}+(\gamma -\gamma _{IR})^{3}\frac{1}{27}k_{3}.
\end{eqnarray}
The subscript $UV$ indicates the free field value, while the other
quantities are evaluated at the scale $\mu $.
We have
\begin{equation}
k_{UV}=-\frac{1}{16}N_{\chi ,}\qquad k_{3\,UV}=\frac{9}{16}N_{\chi
},\qquad
k_{2UV}\equiv -\frac{9}{16}N_{\chi }\frac{N_{c}}{N_{f}},\quad k_{1\,UV}=%
\frac{9}{16}N_{\chi }\left( \frac{N_{c}}{N_{f}}\right) ^{2}.
\end{equation}
Finally, collecting (\ref
{eq:4.6}) and (\ref{gt}), we find the non-perturbative results
\begin{eqnarray}
c&=&c_{UV}+\frac{5}{6}(\gamma -\gamma _{IR})k+\frac{5}{6}\gamma
_{IR}k_{UV}+
\frac{1}{2}h, \nonumber \\
a&=&a_{UV}+\frac{1}{2}(\gamma -\gamma
_{IR})k+\frac{1}{2}
\gamma _{IR}k_{UV}+\frac{1}{2}h.  \label{try}
\end{eqnarray}

Let us compare our results with two-loop calculations. The work of \cite
{jack} shows that $a(g)$ has no 2-loop corrections and in \cite{noi} the
result
\begin{equation}
c(g)=\frac{1}{24}\left( 3N_{v}+N_{\gamma }-\gamma
_{i}^{i}+N_{v}\frac{\beta {%
(g)}}{g}\right)   \label{eq:4.7}
\end{equation}
for $c$ was obtained. We see that the two-loop correction is the sum of a
contribution proportional to $\gamma $ and a contribution
proportional to $\beta (g)/g$.
Formulas (\ref{try}) show that the coefficients of
the $\gamma$ terms are
\begin{eqnarray}
\frac{5}{6}k_{UV}+\frac{1}{2}k_{1\,UV}-\frac{1}{3}\gamma _{IR}k_{2\,UV}+%
\frac{1}{18}\gamma _{IR}^{2}k_{3\,UV} &=&-\frac{1}{24}N_{c}N_{f},
\nonumber \\
\frac{1}{2}k_{UV}+\frac{1}{2}k_{1\,UV}-\frac{1}{3}\gamma _{IR}k_{2\,UV}+%
\frac{1}{18}\gamma _{IR}^{2}k_{3\,UV} &=&0,
\end{eqnarray}
for $c$ and $a$, respectively, in agreement with the two-loop results of
\cite{jack} and (\ref{eq:4.7}).
Note that $\gamma_{i}^{i}=N_{c}N_{f}\gamma $
in the present notation and $\gamma $ is the anomalous dimension of the
gauge invariant $\widetilde{Q}Q$ which is twice the value of the field
anomalous dimension used in \cite{noi}. The comparison of the
coefficient of
the $\beta (g)/g$ terms, instead, is more subtle and requires the precise
knowledge of the two-loop corrections to the functions $k,$
$k_{1},$ $k_{2}$
and $k_{3}$. For example, a two-loop correction to $k$
is expected  from the photon chiral anomaly in an external
gravitational field \cite{dolgov}.

In the infrared limit $\gamma -\gamma _{IR}$ vanishes (it is the
numerator
of $\beta (g)$) and the central charge flows are
\begin{equation}
c_{IR}-c_{UV}=\frac{N_{c}N_{f}}{48}\gamma _{IR}\left(
3\frac{N_{c}}{N_{f}}+9%
\frac{N_{c}^{2}}{N_{f}^{2}}-4\right)
,~~~~a_{IR}-a_{UV}=-\frac{N_{c}N_{f}}{48%
}\gamma _{IR}^{2}\left( 2+3\frac{N_{c}}{N_{f}}\right) .  \label{eq:5.14}
\end{equation}

We can check agreement with the two loop results also from these
formulas,
but only in the weakly coupled region $\gamma _{IR}\ll 1$. Again, the
coefficients of $\gamma _{IR}$ are the expected ones. The fact that
$a$ is
two-loop uncorrected is exhibited by the appearance of $\gamma
_{IR}^{2}$ in
the expression of $a_{IR}-a_{UV}$. The term proportional to $\beta
(g)/g$ in
(\ref{eq:4.7}), on the other hand, cannot be reproduced in the
flows (\ref
{eq:5.14}): although it is nonvanishing at the order $g^{2}$, it is
cancelled at criticality by the higher loop corrections.

\begin{center}
*~~~*~~~*
\end{center}

In the considerations above we have discussed the central
functions along the RG flows.
If one is only interested
in the fixed
point values of  $c$ and $a$ , one can use a
shortcut because the coefficient  $(  1 - 3N_c/N_f -\gamma       )$
vanishes at
the IR fixed
point, so $R_\mu$ and $S_\mu$ effectively coincide there.
 It is again crucial that
the internal anomaly-free current $S_{\mu }$ of (\ref{eq:2.4}) has
one-loop
exact external anomalies, so that the IR\ values of the $R_{\mu
}$-anomalies
coincide with the UV values of the corresponding $S_{\mu
}$-anomalies, which
makes them computable. Thus, for example,
\begin{eqnarray}
\frac{\partial}{\partial x_\mu} \langle R_\mu (x) R_\nu  (y)R_\rho (z)
\rangle_{IR} &=&
\frac{\partial}{\partial x_\mu} \langle S_\mu(x) S_\nu (y)S_\rho (z)
\rangle
_{IR} \nonumber\\
&=& \frac{\partial}{\partial x_\mu} \langle  S_\mu (x)S_\nu (y)S_\rho (z)
\rangle_{UV}
\label{shortcut}
\end{eqnarray}
  and the last anomaly can be obtained from the one-loop graphs of
$S_\mu^{UV} = R_{\mu} + \frac{1}{3}(1-3N_c/N_f) K_\mu$ with $R_\mu$
and $K_\mu$
defined in (\ref{eq:2.2}).
We find
\begin{equation}
5a_{IR}-3c_{IR} = \frac{9}{16}\left(N_c^2 -1
-2N_cN_f\left(\frac{N_c}{N_f} \right)^3
\right)
\label{acir}
\end{equation}
In the UV limit, $\frac{\partial}{\partial x_\mu} \langle R(x)_\mu
R(y)_\nu
R(z)_\rho \rangle$ is determined by the one-loop fermion
triangle graphs of $R_\mu$ in (\ref{eq:2.2}) which yield
\begin{equation}
5a_{UV}-3c_{UV} = \frac{9}{16} \left(N_c^2-1 -
\frac{2N_cN_f}{27}\right).
\label{acuv}
\end{equation}
A similar argument  applied to $\langle TTR \rangle$
can be used to determine the flow of $c-a$. One finds
\begin{eqnarray}
c_{IR}-a_{IR} &=& \frac{1}{16}(N_c^2+1 ) \nonumber\\
c_{UV}-a_{UV} &=& -\frac{1}{16}\left( N_c^2-1-\frac{2}{3}N_fN_c \right)
\label{cairuv}
\end{eqnarray}
(The triangle graph contributions in (\ref{acir}-\ref{cairuv}) are normalized
consistent with (\ref{ca}).)
The flows of $5a-3c$ and $c-a$ agree with the infrared limits of
(\ref{gt})
and (\ref{eq:4.6}), respectively.

To conclude, the IR values of the two gravitational central charges
in the
electric theory are
\begin{equation}
c_{IR}={\frac{1}{16}}\left(
7\,N_{c}^{2}-2-9{\frac{N_{c}^{4}}{N_{f}^{2}}}
\right) ,~~~~a_{IR}={\frac{3}{16}}\left(
2\,N_{c}^{2}-1-3{\frac{N_{c}^{4}}{
N_{f}^{2}}}\right) .  \label{gru}
\end{equation}
These results will be discussed in the next section.
\section{Discussion}

\setcounter{equation}{0}

The techniques developed here are quite generally applicable to any
supersymmetric gauge theory, provided the theory flows to an IR fixed
point and the gauge  $\beta$-function has the general structure of the
NSVZ formula, so that anomalous dimensions can be determined at the fixed
point. Specific applications have been made to the electric N=1 $SU(N_c$)
series in the conformal window $3N_c/2 < N_f <3N_c$ and their
magnetic duals.

It should be noted that our formulas for central charge flows do not give
new tests of duality because the $S_{\mu }$ current whose anomalies
agree in
the electric and magnetic theories \cite{emduality} coincides at the IR
fixed point with the $R_{\mu }$ current for which the anomalies are
various
linear combinations of the IR central charges. We  derive here the
flows $%
c_{IR}-c_{UV}$ and $a_{IR}-a_{UV}$ in the magnetic case by this shortcut.
The IR values are just those given in (\ref{gru}) for the electric
theory,
and the UV values are determined by the free-field content. In this
way we
find, in the magnetic theory,
\begin{eqnarray}
c_{IR}-c_{UV} &=&\frac{1}{24}\left( 1-{\frac{3}{2}}{\frac{N_{c}}{N_{f}}}%
\right) \left( 9\frac{N_{c}^{3}}{N_{f}}-6N_{c}^{2}+6N_{f}^{2}+N_{c}N_{f}%
\right) ,  \nonumber \\
a_{IR}-a_{UV} &=&-\frac{1}{12}\left(
1-{\frac{3}{2}}{\frac{N_{c}}{N_{f}}}%
\right) ^{2}\left( 3N_{c}^{2}+4N_{c}N_{f}+3N_{f}^{3}\right) .
\end{eqnarray}
In these formulas, $\gamma _{q}^{IR}$ is an overall factor and,
again, the
correct two-loop results in the weakly coupled region of the magnetic
theory, $\gamma _{q}^{IR}\ll 1$, are reproduced. Note once again the
appearance of $(\gamma _{q}^{IR})^{2}$ in $a_{IR}-a_{UV}.$

Let us discuss our results for central charge flows from the
viewpoint of $c$%
-theorem expectations. The results (\ref{eq:2.31}) and (\ref{eq:m.9}) for
central charges of flavor currents in the electric and magnetic theories
both satisfy $b_{IR}-b_{UV} >0$ in the entire conformal window, so
there is
no $c$-theorem for flavor.

For the central charge $c$ of the $TT$ OPE and the Euler anomaly
coefficient
$a$, there is earlier work by Bastianelli \cite{bastianelli} for
$SU(N_{c})$
SQCD with $N_{f}=0$ and in the confinement range $N_{c}\leq N_{f}\leq
3N_{c}/2$. The IR limit is a free theory of massless excitations of
independent gauge invariant composite operators. In an empirical
approach,
Bastianelli computed the free field values in the IR and UV and
found both $%
c_{IR}-c_{UV}<0$ and $a_{IR}-a_{UV}<0$, thus suggesting that a
$c$-theorem
holds in supersymmetry for the central charges $c$ and $a$. Our
techniques
permit the extension of these $c$-theorem tests to the conformal window
where there is an interacting IR fixed point.

The results for $c$ show both positive and negative flows. In the
electric
theory $c_{IR}-c_{UV}$ is negative near the lower edge of the conformal
window, but positive near the upper edge. In the magnetic theory $%
c_{IR}-c_{UV}>0$ in the entire range. We conclude that there is no
supersymmetric $c$-theorem for $c$. Discussions about this issue have
already appeared in the literature in the domain of weakly coupled
theories,
both supersymmetric , for example in
\cite{noi} using
formula (\ref{eq:4.7}), and nonsupersymmetric \cite{Cappelli}.

The story of the Euler central charge $a$ is very different,
since we find $a_{IR}-a_{UV}<0$ for both electric and magnetic
theories in
the full range of the conformal window. We believe that this is strong
(because nonperturbative) new evidence in support of an $a$-theorem
(i.e. a  $c$-theorem for the
Euler anomaly) and thus of irreversibility of the RG flow in
quantum field
theory.
This is the first time that these ideas have been tested in strongly
coupled theories.

One can also consider the central charge flow when quarks masses are
introduced or the Higgs mechanism is used to give masses to some
gauge fields and their superpartners. In either case, the IR values
of the
 central charges change,
while the UV values are the same as before. One finds, from (\ref{gru})
and the free field UV values,
\begin{equation}
a_{IR}-a_{UV}=\frac{1}{48}\left( 18 N_c^2-
27\frac{N_c^4}{N_f^2}-9{N_c^{\prime }}^2-2{N_c^{\prime
}}{N_f^{\prime }}\right)
{}.
\end{equation}
Here the IR values $N_{c}$ and $N_{f}$ are assumed to be in the conformal
window, in order to have an IR fixed point, while the UV values $%
N_{c}^{\prime }$ and $N_{f}^{\prime }$ are subject only to the
conditions $%
N_{c}^{\prime }\geq N_{c}$, $N_{f}^{\prime }\geq N_{f}$ and
$N_{f}^{\prime
}\leq 3N_{c}^{\prime }$. The condition $a_{IR}-a_{UV}<0$ is satisfied in
this more general situation, where some fields are integrated out in the
flow.

Let us discuss other positivity conditions suggested by the simple
physical
intuition that the central charges count degrees of freedom, variously
weighted according to the external field one is using (gravitational,
flavor, etc.). In particular, one expects the following inequalities to
hold,
\begin{equation}
c_{IR}\geq 0,~~~~~a_{IR}\geq 0,~~~~~{\frac{\partial
c_{IR}}{\partial N_{f}}}%
\geq 0,~~~~~{\frac{\partial a_{IR}}{\partial N_{f}}}\geq 0,~~~~~{\frac{%
\partial c_{IR}}{\partial N_{c}}}\geq 0,~~~~~{\frac{\partial a_{IR}}{%
\partial N_{c}}}\geq 0.  \label{ineq}
\end{equation}
The first inequality is rigorous, since $c$ appears in the $<TT>$%
-correlator. The second inequality states that $a$ also should
count degrees
of freedom and has been recently proposed in \cite{latorreosborn}.
The other
inequalities express the expectation that adding matter fields (by
varying $%
N_{f}$) or enlarging the color group (by varying $N_{c}$) should increase
the number of degrees of freedom.

The inequalities (\ref{ineq}) obviously hold in the UV and we would
like to
discuss them in the IR. Imposing them on our results (\ref{gru}) we
find the
following restrictions, in the large $N_{c}$ and $N_{f}$ limits,
\begin{equation}
N_{f}^{2}\geq {\frac{9}{7}}N_{c}^{2},~~~~N_{f}^{2}\geq {\frac{3}{2}}%
N_{c}^{2},~~~~{\rm n.r.},~~~~N_{f}^{2}\geq
{\frac{18}{7}}N_{c}^{2},~~~~{\rm %
n.r},~~~~N_{f}^{2}\geq 3N_{c}^{2},
\end{equation}
where ${\rm n.r.}$ means ``no restriction''. The first inequality, as we
said, is rigorous and so puts a limit on the region where the IR
fixed point
exists. This region contains the full conformal window, in agreement with
electric-magnetic duality. We see that below a certain value of
$N_{f}$ our
treatment necessarily breaks down, which means that no IR fixed point
exists. This is expected, since, for example, pure supersymmetric
Yang-Mills
theory ($N_{f}=0$) has no IR fixed point, according to the NSVZ
exact $\beta$-function.

The other inequalities of (\ref{ineq}) have not been proved and so our
discussion is purely speculative. One can observe that all but the
last one
are satisfied in the entire conformal window. So, if one assumes
electric-magnetic duality, then the physical intuition that we
suggested is
not completely correct and there is a region of the conformal
window where
enlarging the gauge group decreases the value of $a$ in the IR. We do not
have a way to resolve this puzzle at the moment, but we believe
that these
remarks are relevant in order to understand the nature of central charges
better.

A final comment concerns the relation between central charges and
't Hooft anomalies.
The 't Hooft anomalies are quantities that are constant at all
energies, so
computable at the free fixed point. They can be regarded as the
invariants
(or ``indices'') of a quantum field theory. The central charges, on the
other hand, have a direct physical meaning and their RG flow is
nontrivial.
In supersymmetric theories the RG interpolation problem can be solved
because at the IR fixed point the central charges are related to the 't
Hooft anomalies. We now show that all the known 't Hooft anomalies can be
expressed in terms of the central charges.

The 't Hooft anomalies that contain at least one vertex $U(1)_S$ are
related by supersymmetry to appropriate terms in the trace anomaly and
therefore to primary central charges.
We know that the $U(1)_S$ anomaly is
proportional to $c-a$, the $U(1)_S^3$ anomaly is proportional to
$5a-3c$ while the
$U(1)_S GG^{\prime }$ anomaly is related to the flavor central
charge $b$,
$G $ and $G^{\prime }$ denoting flavor groups other than $U(1)_S$.
Other $G_{1}G_{2}G~$ 't Hooft anomalies do not contain the vertex
$U(1)_S$.
They
are related to certain secondary central charges $c^{\prime }$.
Using the
construction of Section 3 of \cite{ccfis}, let us consider the
$G$-channel
of the $\langle G_{1}(x)G_{2}(y)G_{3}(z)G_{4}(z) \rangle$ four-point
function, where the limit $|x-y|,|z-w|\ll |x-z|$ is taken.
These secondary
central charges are rather special, since the intermediate channel
is again
a conserved current.
The value of $c^{\prime }$ is simply expressed in terms
of a primary central function $b_{G,G}$ and two 't Hooft anomalies.
More
precisely,
\begin{equation}
c_{G_{i};G}^{\prime }=(G_{1}G_{2}G)\,b_{G,G}\,(G,G_{3},G_{4})
\end{equation}
$(G_{i}G_{j}G)$ denoting the value of the appropriate triangle anomaly.
Therefore the set of primary and (special) secondary central charges
contains the set of 't Hooft anomalies. An implication of this remark is
that the central charges in question coincide in both the electric and
magnetic theories. The secondary central charges that are not included in
this set, on the other hand, do not correspond to known 't Hooft
anomalies,
but they should also coincide in the electric and the magnetic
theory. One
possible extension of the analysis of this paper is the study of
these new
central charges, with the purpose of testing non-abelian
electric-magnetic
duality.

It would be interesting to extend our investigation to nonsupersymmetric
theories. The properties of supersymmetry have been intrinsically used in
our derivation. In particular, crucial roles were played by the exact
expressions for  $\beta$-functions and the  relation between
the trace anomaly and the chiral anomaly of the $R$-current. In view of
this, the nonsupersymmetric generalization is nontrivial.
A relatively simple case to start with is pure Yang-Mills theory.
The central functions $c(\alpha )$ and $a(\alpha )$ should flow
to zero in the IR, since the theory is expected to have a mass gap.

Finally, the other interesting open problem is to rigorously derive the
$a$-theorem (at least in the class of supersymmetric theories) and
definitively prove that the RG flow is irreversible in quantum
field theory.

\section{Acknowledgements}

Two of the authors (D.A. and D.Z.F.) would like to thank CERN for
its kind
hospitality during the final phase of this work. The research of D.A. was
partially supported by EEC grants CHRX-CT93-0340 and TMR-516055. The
research of D.Z.F. was supported by NSF grant PHY-97-22072. The
research of
M.T.G. was supported by NSF grant PHY-96-04587. The research of
A.A.J. was
supported in part by the Packard Foundation and by NSF grant
PHY-92-18167.
\vspace{0.2in}

\appendix

\section{Appendix}
\renewcommand{\theequation}{A.\arabic{equation}}
\setcounter{equation}{0}

\newcommand{\beq}{\begin{equation}}
\newcommand{\eeq}{\end{equation}}
\newcommand{\bea}{\begin{eqnarray}}
\newcommand{\eea}{\end{eqnarray}}
\newcommand{\ena}{\end{eqnarray}}

\renewcommand{\a}{\alpha}
\renewcommand{\b}{\beta}
\renewcommand{\c}{\gamma}
\renewcommand{\d}{\delta}
\newcommand{\q}{\theta}

\newcommand{\pa}{\partial}
\newcommand{\G}{\Gamma}
\renewcommand{\l}{\lambda}
\newcommand{\m}{\mu}
\newcommand{\M}{\Mu}
\newcommand{\n}{\nu}
\newcommand{\N}{\Nu}
\newcommand{\f}{\phi}
\newcommand{\vf}{\varphi}
\newcommand{\F}{\Phi}
\newcommand{\x}{\chi}
\newcommand{\X}{\Chi}
\newcommand{\p}{\pi}
\renewcommand{\P}{\Pi}
\newcommand{\vp}{\varpi}
\newcommand{\R}{\Rho}
\newcommand{\s}{\sigma}
\renewcommand{\S}{\Sigma}
\renewcommand{\t}{\tau}
\newcommand{\T}{\Tau}
\newcommand{\y}{\upsilon}
\newcommand{\Y}{\Upsilon}
\renewcommand{\o}{\omega}
\renewcommand{\O}{\Omega}

\newcommand{\Db}{\bar{D}}
\newcommand{\Wb}{\bar{W}}
\newcommand{\Fb}{\bar{F}}
\newcommand{\Hb}{\bar{H}}
\newcommand{\Pb}{\bar{P}}
\newcommand{\phib}{\bar{\phi}}
\newcommand{\sigmab}{\bar{\sigma}}
\newcommand{\Sigmab}{\bar{\Sigma}}
\newcommand{\Phib}{\bar{\Phi}}
\newcommand{\psib}{\bar{\psi}}
\newcommand{\Psib}{\bar{\Psi}}
\newcommand{\chib}{\bar{\chi}}
\newcommand{\Jb}{\bar{J}}
\newcommand{\pbar}{\bar{p}}
\newcommand{\jmb}{\bar{\jmath}}
\newcommand{\ad}{{\dot{\alpha}}}
\newcommand{\bd}{{\dot{\beta}}}
\newcommand{\gd}{{\dot{\gamma}}}
\newcommand{\dd}{{\dot{\delta}}}

\newcommand{\Del}{\nabla}
\newcommand{\Delb}{\bar{\nabla}}

We present in this Appendix a superspace derivation of the
coefficients $c-a$ and $5a-3c$ appearing in (\ref{eq:4.4}).

For any matter system coupled to  (background) supergravity  the
supercurrent
and supertrace are defined by
\begin{equation}
J_{\a \ad} = \frac{\d \G}{\d H^{\a \ad}} ~~~~,~~~ J = \frac{\d
\G}{\d \phi^3}
\end{equation}
where $\G$ is the (classical or quantum) action, $H^{\a \ad}$ is the
supergravity
vector prepotential superfield, and $\phi$ is the superconformal
(chiral superfield) compensator. The superfield $H^{\a \ad}$  contains
the vierbein and,
as its last, $\theta^2 \bar{\theta}^2$ component, the axial vector
auxiliary field denoted here by $V^{\a \ad}$.
The supercurrent and supertrace  are related by the (local supersymmetry)
 conservation equation
\begin{equation}
\Del^{\ad} J_{\a \ad} = \Del _{\a} J
\end{equation}
If the theory is superconformal the action is independent of the
compensator
and the supertrace vanishes.
Otherwise, $J$ contains the trace anomaly, the supersymmetry
current $\c$-trace
anomaly,
and the chiral anomaly for the $R$-current. The $R$-current
appears as the first component of the supercurrent or,
equivalently, as the
term in the action that couples to  $V^{\a \ad}$.

For  supersymmetric matter systems (scalar or vector multiplets)   the
supertrace
has been computed  in terms of so-called super-$b_4$ coefficients
in refs.
\cite{mcarthur,buch} ( however, we use in this Appendix
the conventions of {\it Superspace}
\cite{superspace}). One finds
\begin{equation}
J= \frac{1}{24\pi^2}[ cW^2 - a \Xi_c]
\end{equation}
where the numerical coefficients $c$, $a$ take the values
\begin{eqnarray}
{\rm Scalar ~multiplet} &:& ~~c= \frac{1}{24} ~~~,~~~~ a= \frac{1}{48}
\nonumber\\
{\rm Vector~ multiplet} &:&~~ c= \frac{3}{24} ~~~,~~~~ a= \frac{9}{48}
\end{eqnarray}
Here $W^2 = \frac{1}{2} W_{\a \b \c}W^{\a \b \c}$ is the square of the
superWeyl
tensor, while
\begin{equation}
\Xi_c =   W^2  + (\Delb^2 +R) (G^2+2 \bar{R}R)
\end{equation}
is the chirally projected superEuler density. More precisely
\begin{eqnarray}
\cal{E} &=& \frac{1}{(4\pi )^2}\left[\int d^4x d^2 \theta \phi^3  \Xi_c
+\int d^4x d^2 \bar{\theta} \phib^3 \bar{  \Xi}_c \right]\\
\cal{P} &=& \frac{1}{(4\pi )^2}\left[\int d^4x d^2 \theta \phi^3  \Xi_c
-\int d^4x d^2 \bar{\theta} \phib^3 \bar{  \Xi}_c\right] =
\frac{1}{(4\pi )^2}\left[\int d^4x d^2 \theta \phi^3 W^2
-\int d^4x d^2 \bar{\theta} \phib^3 \bar{ W}^2\right]  \nonumber
\end{eqnarray}
give the Euler number and Pontrjagin number, respectively.
The chiral  superfields  $W_{\a \b \c}$ and
$R$, and the real superfield
$G_{\a \ad}$
are the three superspace curvatures.
 In our conventions
$G^2 =\frac{1}{2} G^{\a \ad}G_{\a \ad}$.

We are interested in the component trace and $R$-current anomalies
for the
matter
system in a background gravity and  axial-vector auxiliary field
 $R$-current source. The trace of the stress-tensor
is obtained from the $\q^\a \bar{\q}^\ad$ component of the supercurrent:
\begin{equation}
\Theta =\frac{3}{8} [\Del^\a , \Delb^\ad ] J_{\a  \ad} |_{\q=0}
=\frac{3}{4}(\Del^2J +\Delb^2 \bar{J}) |_{\q =0}
\end{equation}
while the divergence of the $R$-current is obtained from the first
component of
the
supercurrent:
\begin{equation}
i \Del^{a } R_a =\frac{1}{2} \{\Del^\a , \Delb^\ad \} J_{\a \ad}
|_{\q =0}=(
\Del^2J - \Delb^2 \bar{J})
|_{\q =0}
\end{equation}
One can obtain the corresponding component anomalies by brute force
$\theta$-expansion
or, as we shall do here, by  definining components
by projection and exploiting the Bianchi identities of the theory.
In doing so,
we will obtain
component results expressed in terms of the component curvature tensor
$R_{abcd}$ and the
axial vector field strength, $V_{ab}$. However, one further step is
necessary:
the standard
supergravity constraints \cite{superspace} lead to component  covariant
derivatives
with nonzero torsion, due to both the gravitino fields (that we set
to zero
here), and to the
axial vector auxiliary field.  Before reading off the component
anomalies, we
must
separate the curvature in terms of the ordinary Einstein curvature and
additional terms,
again proportional to the field strength  $V_{ab}$.

We obtain the component results as follows:  the $W^2$ term in the
supertrace
leads to
\begin{equation}
\frac{1}{2}\Del^2 W^2 \pm  \frac{1}{2}\Delb^2 \bar{W^2} =
\frac{1}{4} \Del^\d
W^{\a \b \c} \Del_\d W_{\a \b \c}
- \frac{1}{4} W_{\a \b \c} \Del^2 W^{\a \b \c} \pm h.c.
\end{equation}
evaluated at $\q =0$, the $\pm$ sign corresponding to either the
trace or the
chiral
anomaly.
Since we are restricting ourselves to a bosonic background the
second term
(proportional at $\theta=0$ to the gravitino field strength) may be
dropped.
We
write each
factor in the first term as
the sum of symmetrized and antisymmetrized (in $\d$ and the already
symmetrized
$\a \b \c$) indices, i.e.
\begin{equation}
\Del_\d W_{\alpha \beta \gamma} = \frac{1}{6}[ \Del_{(\delta}
W_{\alpha \beta \gamma) } + C_{ \alpha
\delta}\Del^{\lambda} W_{\lambda  \beta \gamma } +
C_{ \beta  \delta}\Del^{\lambda} W_{\lambda  \gamma \alpha }
+ C_{ \gamma   \delta}\Del^{\lambda} W_{\lambda  \alpha \beta }]
\end{equation}
For the last three terms we use the Bianchi identity $\Del^{\l}
W_{\l \b \c }
=\frac{i}{2}\Del_{(\b}^{~~\bd} G_{\c ) \bd}$. Evaluating at $\q=0$, the
(totally symmetrized)
first term gives the self-dual part (in spinor notation) of the
component Weyl
tensor,
$W_{\a \b \c
\d} = \frac{1}{4!}  \Del_{(\delta}  W_{\alpha \beta \gamma) } $
 (but we emphasize again that it still contains some torsion pieces).
The second term, using the definition of the axial vector auxiliary
field  $V_a
= G_a|_{\theta =0}$
leads to the self-dual part of its field strength, $V_{\b \c}=
\frac{1}{2}
\partial_{(\b \bd}
V_{\c)}^{~\bd}$.

For the $G^2$ terms in the superEuler density, dropping irrelevant
terms, we
have
\begin{equation}
(\Del^2 \Delb^2 \pm \Delb^2 \Del^2) G^2 = - \frac{1}{2} [ \Del^\a
\Delb^\ad
G^{\b \bd}
 \Del_\a \Delb_\ad G_{\b \bd} \pm
 \Delb^\ad \Del^\a G^{\b \bd} \Delb_\ad \Del_\a G_{\b \bd} ]
\end{equation}
With suitable (anti)symmetrization and use of the Bianchi identity
$\Del^\ad
G_{\a \ad} = \Del_\a R$, one has, for example,
\begin{equation}
\Delb_\ad \Del_\a G_{\b \bd} = \frac{1}{4} \Delb_{(\ad}\Del_{(\a}
G_{\b ) \bd
)}
+\frac{1}{4}C_{\bd \ad} \Del^{~~\gd}_{ (\a}G_{\b) \gd } + \frac{1}{2}
C_{\bd \ad}
C_{\b \a} \Delb^2 \bar{R}
\end{equation}
At $\theta =0$  the middle term
is
again  expressible in terms of the selfdual part of  $V_{ab}$.
 The first  and last terms can be shown  to be
expressible in terms of the (torsionful) Ricci tensor and scalar.
as follows: we have
the identity expressing the component curvature tensor
$R_{abcd} = C_{\dot{\gamma} \dot{\delta}} R_{ab \gamma \delta} +
C_{\gamma \delta}R_{ab\dot{\gamma} \dot{\delta}}$  in terms of
superspace
objects  \cite{superspace,buchbinder}
\begin{eqnarray}
R_{ab \c \d}& \equiv & R_{ \a \ad , \b \bd , \c \d }  =\frac{1}{6}
C_{\ad \bd}
\Del_{(\a}W_{\b \c \d )} - \frac{1}{4} C_{\a \b}
\Del_{(\ad}\Del_{(\c} G_{\d )\bd )} \nonumber\\
&-&\frac{i}{12}C_{\ad \bd} [C_{\b \c} \Del_{(\a}^{~~ \gd}G_{\d )\gd}
+C_{\b \d} \Del_{(\a}^{~~\gd}G_{\c )\gd}] +\frac{1}{2}C_{\ad \bd} C_{\b
(\c}C_{\d ) \a}
(\Delb^2 \bar{R}+2R\bar{R})
\end{eqnarray}
where the right hand side is evaluated at $\theta =0$. In particular
\begin{equation}
\Delb_{(\ad}\Del_{( \c}G_{\d )\bd} |_{\theta =0}= 2 R^{\b}_{ ~(\ad , \b
\bd ), \c \d}
\end{equation}
and, similarly
\begin{equation}
\Delb^2 \bar{R} +2 R \bar{R}|_{\theta = 0}=- \frac{1}{6}
R^{\c \ad ~\d}_{~~~,~\ad , \c \d}
\end{equation}
Our final ingredient expresses the component torsion  and  the torsionful
connection in terms of the axial vector auxiliary field and the
torsionless connection  \cite{superspace,buchbinder}:
\begin{eqnarray}
T_{ab}^{~~c} &\equiv& T_{\a \ad , \b \bd}^{~~~~\c \gd} =
i\left( C_{\a \b} \d_{\ad}^{~\gd}V^\c_{~\bd} - C_{\ad \bd} \d_\a^{~\c}
V_\b^{~\gd}\right)\nonumber\\
\omega_{abc}&=& \omega_{abc}(e) - \frac{1}{2} \varepsilon_{abcd}V^d
\end{eqnarray}
This allows rewriting the component  torsionful curvatures in terms of
the ordinary curvature tensor and additional $V_{ab}$ terms.

Assembling all the ingredients,
we finally obtain the following contributions
to the trace and $R$-current anomalies:
\begin{eqnarray}
\Theta &=&\frac{c}{32 \pi^2}[ (W_{\a \b \c \d})^2 +(\bar{W}_{\ad
\bd \gd
\dd})^2]
+ \frac{c}{3 \pi^2}[(V_{\a \b})^2 +(\bar{V}_{\ad \bd})^2 ] \nonumber\\
&&-\frac{a}{32\pi^2}[ (W_{\a \b \c \d})^2 +(\bar{W}_{\ad \bd \gd \dd})^2
-2 R^{\a \b \ad \bd}R_{\a \b \ad \bd} +\frac{8}{3}(R_{\a \b}^{~~\a \b})^2]
\nonumber\\
\partial^a R_a &=&\frac{ c-a}{24 \pi^2}[(W_{\a \b \c \d})^2  -
(\bar{W}_{\ad
\bd \gd \dd})^2]
-4\frac{5a-3c}{9 \pi^2} [ (V_{\a \b})^2 - (\bar{V}_{\ad \bd})^2 ]
\end{eqnarray}
in terms of selfdual and anti-selfdual parts of the curvature
tensors and the
axial vector field strength. This can  be rewritten in the form given in
(\ref{eq:4.2.5}) and (\ref{eq:4.4}).

\end{document}